\documentclass[fleqn,10pt]{wlscirep}
\usepackage[utf8]{inputenc}
\usepackage[T1]{fontenc}
\usepackage{amsmath}
\usepackage{amsfonts}
\usepackage{amssymb}
\usepackage{graphicx}
\usepackage{empheq}

\newcommand{\partialdiff}[2]{\frac{\partial{#1}}{\partial{#2}}}
\newcommand{\txtpow}[1]{{\mbox{\scriptsize{#1}}}}

\title{Precision and informational limits in inelastic optical spectroscopy}

\author[1,2]{Peter T\"or\"ok}
\author[2,*]{Matthew R. Foreman}
\affil[1]{School of Physical and Mathematical Sciences, Nanyang Technological University, Singapore}
\affil[2]{Blackett Laboratory, Department of Physics, Imperial College London, Prince Consort Road, London, SW7 2AZ, UK }

\affil[*]{matthew.foreman@imperial.ac.uk}

\begin{abstract}
Using Fisher information and the Cram\'er-Rao lower bound, we analyse fundamental precision limits in the determination of spectral parameters in inelastic optical scattering. General analytic formulae are derived which account  for the instrument response functions of the dispersive element and relay optics found in practical Raman and Brillouin spectrometers. Limiting cases of dispersion and diffraction limited spectrometers, corresponding to measurement of Lorentzian and Voigt lineshapes respectively, are discussed in detail allowing optimal configurations to be identified. Effects of defocus, spherical aberration, detector pixelation and a finite detector size are also considered.  
\end{abstract}
\begin{document}

\flushbottom
\maketitle

\thispagestyle{empty}

\section*{Introduction}
Interactions between either acoustic or optical phonons in a material and an incident photon, can give rise to inelastic scattering, known more specifically as Brillouin or Raman scattering respectively \cite{Brillouin1922,Raman1927}. Such processes provide a means by which to probe the vibrational, micro-mechanical and compositional properties of samples \cite{Smith2005a,Chowdhury1998,Koski2013,Sheng2017,Wu2018}. As such, recent decades have seen development of a wealth of experimental techniques for acquisition of inelastic scattering spectra including fibre, imaging and near field based setups \cite{Kabakova2017,Zavaleta2013,Koski2005,Scarcelli2008,Jahncke1995,Steidtner2008}. Inelastic scattering is, however, an intrinsically weak process. Despite improvements in achievable  signal to noise ratios using surface enhancements, stimulated processes, interferometric or heterodyne detection \cite{Vohringer1995,Schlucker2011,Ballmann2017,Tanaka1995,Antonacci2015}, the value of experimental data can be dramatically degraded by noise. Systematic quantitative analysis of achievable precision in inelastic optical spectroscopy is, however, hitherto lacking and will thus form the focus of this article. Evaluation of such limits is not only of importance in terms of aiding system design \cite{Bowie2000,Antonacci2013,Coker2018}, particularly in scenarios with limited photon budgets, but can also enable benchmarking of data processing algorithms and analysis protocols \cite{Craggs1996,Xiang2018}.

Within this context, we first detail the model of signal acquisition adopted throughout this work. We restrict attention to measurement of spontaneous scattering spectra by means of angularly/spatially dispersive spectrometers, as opposed to scanning etalon based alternatives. In particular, our treatment incorporates the imperfect nature of the dispersive element required in any spectroscopic experiment, and allows for additional, potentially aberrated, relay optics. We then proceed to outline the information theoretic precision limit, given by the Cram\'er-Rao lower bound, as applied  to the problem of extracting spectral parameters in inelastic spectroscopy. Finally, we apply these results to a number of limiting cases and numerical examples.

\section*{Detection model in spectrometers}\label{sec:detection}
An ideal spontaneous or thermally excited Brillouin or Raman spectrum consists of a central Rayleigh scattering peak flanked by the Stokes and anti-Stokes  scattering peaks corresponding to generation or annihilation of a phonon. If the Rayleigh peak is centred at  frequency $\omega_0$, then the two inelastic peaks are centred at $\omega_{\pm 1} = \omega_0 \pm \Omega$, where $\Omega$ is the inelastic frequency shift. Assuming each peak has an arbitrary lineshape $F(\omega,\omega_p,\Gamma_p)$ for $p = 0,\pm 1$, the associated complex amplitude of each spectral frequency component, denoted $\omega$, is given by
\begin{align}
s(\omega) =  \sum_{p = -1}^{+1} A_p   F(\omega;\omega_p,\Gamma_p)\label{eq:Lorentz_spec}
\end{align}
where $\Gamma_{-1}= \Gamma_{+1} =\Gamma$ is the intrinsic full-width half-maximum (FWHM) of the inelastic peaks, $\Gamma_0$ is the intrinsic FWHM of the Rayleigh peak and $A_{\pm 1,0}$ describes their respective magnitudes. The lineshape is assumed to be normalised such that $\int_{-\infty}^{\infty} | F(\omega)|^2 d\omega = 1$, meaning that $|A_p|^2$ describes the total power in a given spectral peak.
Assuming any measurements are made over a time long compared to the phonon coherence time and restricting attention to spontaneous inelastic processes it follows that	the power spectrum is
\begin{align}
S(\omega) = |s(\omega)|^2 = \sum_{p = -1}^{+1} |A_p|^2  |F(\omega;\omega_p,\Gamma_p)|^2. \label{eq:Ispec_omega}
\end{align}
In this work we will assume that the spectral lineshapes are Lorentzian in nature, i.e. $F(\omega,\omega_p,\Gamma_p) =L(\omega;\omega_p,\Gamma_p)$ where 
\begin{align}
L(\omega;\omega_p,\Gamma_p)&=\sqrt{\frac{2}{\pi\Gamma_p}}  \frac{i \Gamma_p/2}{(\omega-\omega_p) + i \Gamma_p / 2} \label{eq:Lorentzian_om}.
\end{align}

In an experimental scenario an observed spectrum is subject to the instrumental response of the spectrometer used to measure it \cite{Wilksch1985,Xiao2004}, such that the measured lineshape may not be a pure Lorentzian.  The core function of an optical spectrometer is to spatially separate each individual frequency component,  $\omega$, present in an input wave. To achieve this a dispersive element, such as a grating or virtually imaged phase array (VIPA) \cite{Shirasaki1999a}, is used to discriminate different frequency components through the angle at which they are diffracted or transmitted. By placing a detector in the far field region of the dispersive element, the angular discrimination is converted to a spatial separation of differing frequency components through diffraction alone. In practice, the dispersive element produces a spatially extended distribution on the detector even for a monochromatic input as can be described by the associated complex amplitude response function $h_{\txtpow{disp}}(x,\omega)$, where $x$ describes the spatial coordinate in the detector plane. Note that since single or stand alone dispersive devices only produce a one-dimensional  angular separation we restrict our discussion to a single spatial coordinate $x$. Nevertheless, our analysis applies to any number of dispersive spectrometers arranged in a crossed configuration along the direction of dispersion. High finesse VIPAs, for example, give rise to Lorentzian peaks on the detector for a monochromatic input. For simplicity, in this work we also assume a linear mapping between the input frequency of light and the ideal location of the resulting peak on the detector, such that 
\begin{align}
h_{\txtpow{disp}}(x,\omega) = h_{\txtpow{disp}}(x - x_0 (\omega)) \label{eq:hdisp1} 
\end{align}
where $x_0(\omega) = \alpha (\omega - \omega_{\txtpow{off}})$, $\alpha$ describes the scaling constant between frequency and real space and $\omega_{\txtpow{off}}\textsl{}$ allows for an arbitrary spectral offset. To account for the finite free spectral range (FSR), $\Omega_{\txtpow{FSR}}$, inherent in a spectrometer we let
\begin{align}
h_{\txtpow{disp}}(x-x_0) = \sum_{q=-\infty}^{\infty} B_q \, \bar{h}_{\txtpow{disp}}(x-x_0 + \alpha q \Omega_{\txtpow{FSR}})  \label{eq:hdispFSR}
\end{align}
where the different FSRs are indexed by $q$ and we have assumed that the lineshape of an individual peak, $\bar{h}_{\txtpow{disp}}$, is the same for each FSR up to a slowly varying amplitude variation described by the factor $B_q$.  We shall also make the further simplifying assumption that $\sum_{q=-\infty}^{\infty} |B_q|^2 = 1$, which assuming $\Omega_{\txtpow{FSR}} \gg \gamma_{\txtpow{disp}}$, where $\gamma_{\txtpow{disp}}$ parametrises the width of $\bar{h}_{\txtpow{disp}}$, implies $\int_{-\infty}^{\infty} |{h}_{\txtpow{disp}}|^2 dx = \int_{-\infty}^{\infty} |\bar{h}_{\txtpow{disp}}|^2 dx$.

Although the angular dispersion of a dispersive element can be converted to a spatial separation by detection in the Fraunhofer zone as discussed above, more commonly additional optics are placed after the dispersive element since this enables smaller device footprints and greater ease of use. One simple means by which this transformation can be achieved is to place a single lens a focal length from the exit surface of the dispersive element so as to achieve a Fourier transform of the output waveform. Introduction of such optics, which are  not ideal, degrades the final detected lineshape for a monochromatic input further. Non-ideality arises not only from the finite numerical aperture of practical optical elements, but also due to aberrations that may be present.

Whilst all lens aberrations modify the shape of the amplitude response function (also known as the point spread function or PSF), the resulting consequences can differ in the context of inelastic spectroscopy. Aberrations that change the location of the maximum of a peak on the detector, for example, can result in erroneous estimates of the absolute frequency of that peak if they are not properly accounted for. Tilt, for example, produces a uniform shift of all peaks which can give rise to a systematic frequency shift, although practically tilt is of little interest as it is easily removable via calibration. Field-dependent shifts, on the other hand, arising from say coma or distortion, not only can give rise to errors in the estimated spectral frequency but also consequently in the inferred Brillouin or Raman shifts. Aberrations that change the observed shape (but not position) of a  peak uniformly throughout the detection field (i.e. defocus and spherical aberration) can instead lead to systematic errors in the determination of the peak width which is a measure of the phonon decay lifetime. Such errors, can however become non-systematic, if an aberration affects the peak width differently depending on field position (e.g. curvature of field).

In addition to their effect on the observed lineshape, strong aberrations can also produce a non-linear spatial dispersion. Minimisation of aberrations is however usually sought through appropriate optical design, such that we can reasonably restrict attention to the weakly aberrated regime. The spectral amplitude response function of the dispersive element and additional optics combined can in this case be described by
\begin{align}
h_{\txtpow{spec}}(x,\omega) = h_{\txtpow{disp}}(x- x_0 (\omega)) \otimes h_{\txtpow{opt}}(x,\omega), \label{eq:spectral_response}
\end{align}
where $h_{\txtpow{opt}}(x,\omega)$ is the PSF of the relay optics and we have implicitly assumed that the response functions are shift invariant. 
Since the spectral bandwidths encountered in Brillouin spectroscopy are small we shall also assume that we can safely neglect chromatic aberrations such that $h_{\txtpow{opt}}(x,\omega) = h_{\txtpow{opt}}(x)$. Within the context of Raman spectroscopy in which larger bandwidths are encountered we  assume that optics are corrected for chromatic aberration.  Although introduced in relation to $h_{\txtpow{disp}}$, the parameter $\alpha$ appearing in $x_0(\omega)$ is now also used to account for additional linear scale factors between $\omega$ and $x$ that may arise from the relay optics. 

\begin{figure}[t!]
	\begin{center}
		\includegraphics[width=0.5\columnwidth]{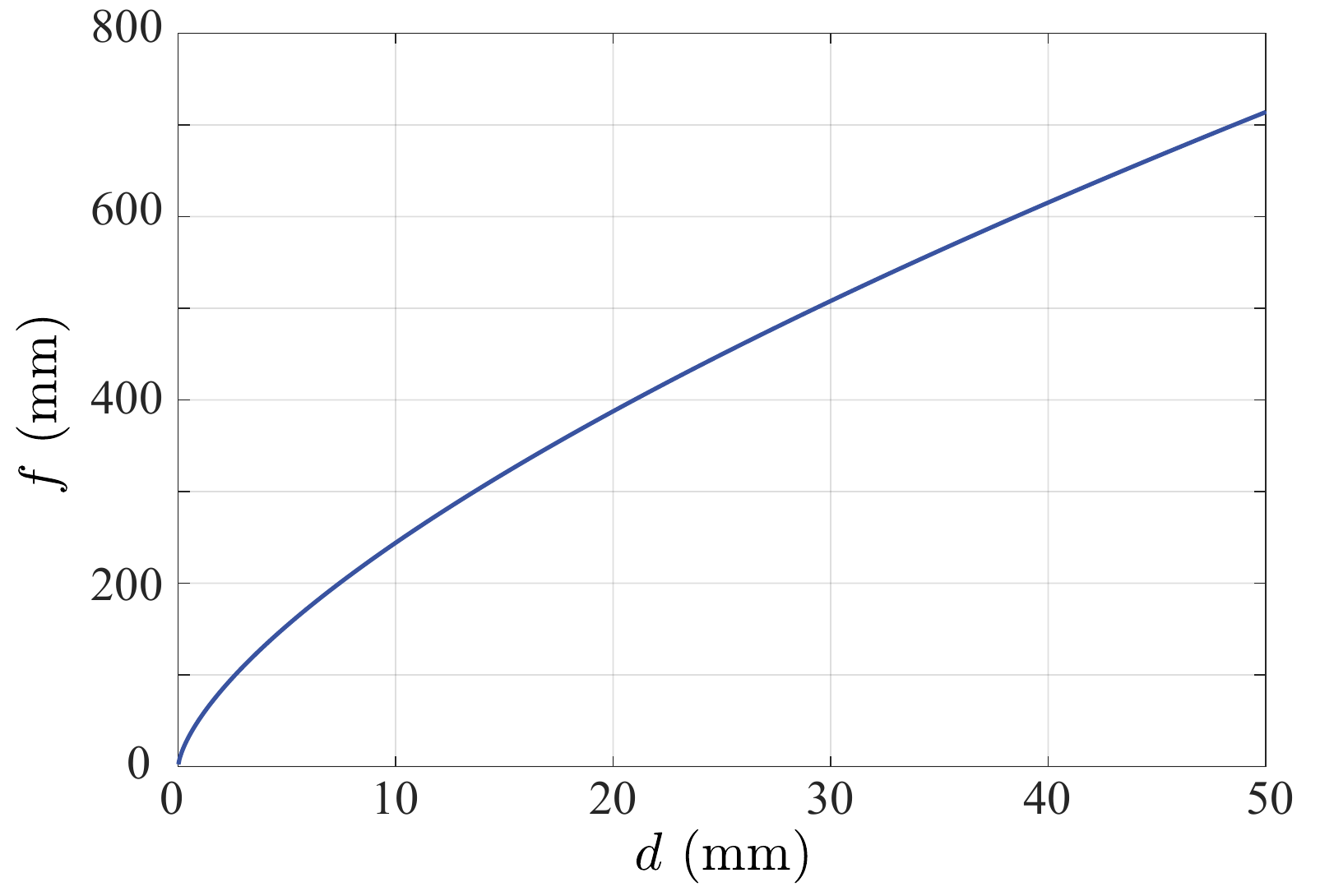}
		\caption{Minimum focal length for which coma in a singlet lens can be safely neglected as determined using the Mar\'echal condition.\label{fig:coma_tol}}
	\end{center}
\end{figure}

The assumption of shift invariance made in Eq.~\eqref{eq:spectral_response} is a limitation to our theory, since it means only field-independent monochromatic aberrations, i.e.~defocus and spherical aberration can be described. In order to gauge under what experimental conditions our results might therefore be applicable, we briefly digress to consider the Mar\'echal condition as applied to possible coma (the lowest order field dependent aberration) in the lens. The Mar\'echal condition \cite{Born1980a} requires the Seidel sum for coma to be less than $1.2\lambda$ for the optical system to be diffraction limited. For a singlet lens, the focal length thus needs to satisfy $f^3 \ge (X^2d^2)/(2.4\lambda)$  where  $d$ is the marginal ray height at the lens. Figure~\ref{fig:coma_tol} shows the minimum required focal length for a singlet assuming a detector size of $X=14$~mm and a wavelength of $\lambda = 561$~nm. For more complex lenses, such as achromatic doublets, a simple analogous criterion does not exist meaning aberration tolerances of these lenses must instead be determined individually. As an example, assuming the same detector size as above, the Seidel sum of an achromatic doublet supplied by a popular optical component manufacturer of focal length $f=100$~mm and diameter $2d=50$~mm, is $S_{II}=1.9\lambda$. Since coma is a linear function of field position, a maximum detector size that can hence be used with this lens is $X=8.8$~mm, thus highlighting the need for careful analysis of aberrations before a lens can be safely used. 

Returning now to our detection model, we note 
that the signal observed on a position resolving detector (initially neglecting pixelation) when a spectrum of  frequencies is input is  proportional to the total incident intensity. Assuming the detector integration time is long relative to the optical periods involved, the observed signal at a position $x$ is hence 
\begin{align}
I_{\txtpow{det}}(x) =  N \Pi\left(\frac{x}{X}\right)\int_0^\infty S(\omega) |h_{\txtpow{spec}}(x,\omega) |^2 d\omega \label{eq:Idet}
\end{align}
where $\Pi(z)$ is the top hat function defined by
\begin{align}
\Pi(z) = \left\{\begin{array}{cc} 
0  & \mbox{for } |z| > 1/2 \\ 
1 & \mbox{for }  |z| < 1/2 \\
1/2 & \mbox{for } |z| = 1/2 \end{array}\right. ,
\end{align} 
$X$ is the full spatial width of the detector, and  $N$ is a normalisation constant which will be discussed later. Using Eq.~\eqref{eq:Ispec_omega} and Eqs.~\eqref{eq:hdispFSR}--\eqref{eq:Idet} we thus have
\begin{align}
&I_{\txtpow{det}}(x) 
=N\Pi\left(\frac{x}{X}\right) \sum_{p,q} |A_{p}|^2|B_{q}|^2 \int_0^\infty\!
|L(\omega;\omega_p,\Gamma_p)|^2
\left|\int_{-\infty}^{\infty}  \bar{h}_{\txtpow{disp}}(x'- x_0(\omega))  h_{\txtpow{opt}}(x' - x ) dx'\right|^2  d\omega. \label{eq:Idet_convnew}
\end{align}

In this work we adopt simple models for the response function of the dispersive element and relay optics. Firstly we shall assume that the lineshape produced by the dispersive element is Lorentzian in profile such that 
\begin{align}
\bar{h}_{\txtpow{disp}}(x) = A L(x,x_0,\gamma_{\txtpow{disp}}) e^{i \kappa x},
\end{align}
where $A$ is a scale factor and we have allowed for a phase tilt for reasons that will become apparent below. The associated coherent transfer function of the dispersive element, defined as the Fourier transform of the amplitude response function, is hence given by the one sided exponential, viz.
\begin{align}
&\widetilde{h}_{\txtpow{disp}}(k_x) = \frac{1}{2\pi}\int_{-\infty}^{\infty} h_{\txtpow{disp}}(x)  e^{-ik_x x} dk_x,  =  \exp\left[\left(-\frac{\gamma_{\txtpow{disp}}}{2} + i x_0\right)(k_x-\kappa)\right]  \Theta[k_x - \kappa], \label{eq:htildedef}
\end{align}
where $k_x$ denotes the spatial frequency coordinate in the Fourier domain and the Heaviside function $\Theta[k_x - \kappa]$ is unity for $k_x \geq \kappa$ and zero otherwise. To ensure the dispersive element is passive ($|\widetilde{h}_{\txtpow{disp}}(k_x)| \leq 1$ for all $k_x$) the scale factor has been set to $A = [2{\pi/\gamma_{\txtpow{disp}}}]^{1/2} $.

To model the PSF $h_{\txtpow{opt}}(x)$ of the relay optics, we consider a single, potentially aberrated, lens of finite numerical aperture, $\text{NA}$, such that \cite{Goodman1996a}
\begin{align}
h_{\txtpow{opt}}(x) = \int_{-\infty}^\infty \Pi\left[\frac{k_x}{k_0\mbox{NA}}\right] \exp\left[i\,W\left(\frac{k_x}{k_0\mbox{NA}}\right)\right] e^{i k_x x} dk_x
\end{align}
where $\widetilde{h}_{\txtpow{opt}}(k_x) = \Pi\left[{k_x}/({k_0\mbox{NA}})\right] \exp[iW(k_x/(k_0 \mbox{NA}))]$ (defined analogously to Eq.~\eqref{eq:htildedef}) is the coherent transfer function of the lens and $W(k_x/(k_0 \mbox{NA}))$ describes a phase aberration in the pupil. In two-dimensional optical systems it is common to represent a perturbed wavefront using the Zernike functions $Z_{nm}(r,\phi)$ because they form an orthonormal complete basis over the domain of the (circular) pupil\cite{Born1980a}, however, for the one-dimensional model employed in this work such a description is unsuitable. Instead, noting that the Legendre polynomials form an orthonormal basis on the interval $[-1,1]$, a more appropriate formalism for the one-dimensional problem\cite{Barakat1965a} is to let 
\begin{align}
W\left(\frac{k_x}{k_0\mbox{NA}}\right) = \sum_{n} a_n L_n\left(\frac{k_x}{k_0\mbox{NA}}\right),\label{eq:Legendreexpansion}
\end{align}
where $L_n(z)$ is the $n$th order Legendre polynomial. Notably, the Legendre polynomial of order $n$ has a similar functional form to the rotationally symmetric Zernike functions, i.e. $Z_{n0}$, and thus a correspondence can be made between the one and two dimensional aberrations. For example, defocus in a two-dimensional system is described by $Z_{20}(r,\phi) = 2r^2 - 1$, whilst the second order Legendre polynomial is given by $L_2(z) = (3z^2 - 1)/2$. Although the numerical value of the coefficients differ between the two polynomials (due to the differing orthogonality domains), both exhibit a quadratic phase perturbation to the wavefront. Accordingly, in a one-dimensional system $L_2$ can be associated with defocus (and similarly for higher order aberrations). With regards to Eq.~\eqref{eq:Legendreexpansion}, it is also important to observe that in the presence of field dependent aberrations $a_n$ is a function of observation position $x$. Our assumption of shift invariance, however, precludes this possibility such that we can only legitimately consider defocus ($n=2$) and spherical aberration. Restricting to primary spherical aberration ($n=4$) only we have $a_n =0$ for $n \neq 2,4$.

Finally we must define the normalisation constant $N$. Specifically, $N$ is defined in terms of the total integrated intensity that would impinge on an infinite detector, $I_\infty =\lim_{X\rightarrow \infty} \int_{-\infty}^{\infty} I_{\txtpow{det}}(x) dx$, such that 
\begin{align}
&I_\infty = N \sum_{p,q} |A_{p}|^2|B_{q}|^2 \int_{-\infty}^{\infty}
\int_0^\infty 
|L(\omega;\omega_p,\Gamma_p)|^2 
\left|h_{\txtpow{spec}}(x - x_0(\omega) ) \right|^2 d\omega dx.
\end{align}
Upon reordering the integrals  we find 
\begin{align}
N = \frac{I_\infty}{\sum_{p=-1}^{+1} |A_{p}|^2}
\frac{1}{ \int_{-\infty}^{\infty} \left|h_{\txtpow{spec}}(x ) \right|^2dx  } . \label{eq:Ndef}
\end{align}
We can evaluate the integral factor in Eq.~\eqref{eq:Ndef} using Parseval's theorem ultimately yielding
\begin{align}
N =\frac{1}{2\pi}\frac{\gamma_{\txtpow{disp}} }{1-\exp[-(k_0 \text{NA}-\kappa) \gamma_{\txtpow{disp}}]}  \frac{I_\infty}{\sum_{p=-1}^{+1} |A_{p}|^2}.
\end{align}
Experimentally it is better to maximise the energy throughput of the spectrometer, which is limited due to the finite numerical aperture of the relay optics. Practically this implies that the  field distribution at the exit surface of the dispersive element (corresponding to $\widetilde{h}_{\txtpow{disp}}$) is positioned so as to maximally fill the entrance pupil of the relay optics (described by the top hat function in $\widetilde{h}_{\txtpow{opt}}$). This corresponds, for the 1D case considered here, to selecting $\kappa = -k_0\text{NA}$ yielding
\begin{align}
N =\frac{1}{2\pi}\frac{\gamma_{\txtpow{disp}}}{1-\exp[-2k_0 \text{NA} \gamma_{\txtpow{disp}}]}  \frac{I_\infty}{\sum_{p=-1}^{+1} |A_{p}|^2}.
\end{align}

\section*{Parametrising precision with Fisher information}\label{sec:FI}
Any experimental study is at heart an estimation problem, in which an observer estimates the value of some parameter(s) of interest from a noise corrupted signal.
Ultimately we wish to consider the accuracy to which the inelastic shift $\Omega$ can be determined since this is the principle parameter of interest in inelastic spectroscopy. It is frequently reasonable to assume that the Rayleigh frequency $\omega_0$ is known \emph{a priori} since measurements are generally performed using a pre-calibrated laser excitation source. For completeness, however, we assume $\omega_0$ is also unknown. Analysis of the precision limits can be simplified by considering the related problem of estimating the centre of mass frequency of the spectrum (determined from a single FSR) $\bar{\omega}$ and the separation of the two inelastic peaks $\bar{\Omega} = 2\Omega$, i.e. we consider estimation of the parameter vector ${\mathbf{w}} = (\bar{\omega}, \bar{\Omega})$. This alternative parametrisation is notably more suitable for cases in which the Rayleigh peak is saturated on the detector, or suppressed \cite{Antonacci2015} which can hence hamper direct determination of $\omega_0$ and $\Omega$. The centre of mass of the spectrum is given by
\begin{align}
\bar{\omega} &= \frac{\int_{-\infty}^{\infty} \omega S(\omega)d\omega}{\int_{-\infty}^{\infty}S(\omega)d\omega} =  \omega_0 + \frac{(|A_{+1}|^2 -|A_{-1}|^2 ) \Omega}{|A_{-1}|^2  +|A_0|^2  +|A_{+1}|^2  } \triangleq \omega_0 + \chi \frac{\bar{\Omega}}{2}.
\end{align}
It hence follows that $\omega_{\pm 1} = \bar{\omega}  - (\chi  \mp  1)\bar{\Omega}/2$. Henceforth we assume that $x_0 =\alpha (\omega - \bar{\omega} + q \Omega_{\txtpow{FSR}})  +\bar{x}$ such that the centre of mass of the spectrum is ideally located at $\bar{x}$. 

In some applications, it may be of interest to estimate further spectral parameters (or indeed required by fitting algorithms), such as the width of each peak from which phonon lifetimes can be determined. For completeness in what follows we shall thus consider the expanded parameter vector $\mathbf{w} = (\bar{\omega}, \bar{\Omega},\{\Gamma_p\},\{|A_{p}|^2\})$, however in our numerical examples we shall principally restrict attention to the achievable precision when determining the inelastic frequency shift. 

The parameter vector $\mathbf{w}$ is not measured directly, but is instead inferred from noisy intensity measurements on the detector. In the previous section we neglected pixelation of the detector, however, in reality this is unreasonable and we thus relax this restriction now. In the ideal noise free case, the measured intensities from each pixel form a data vector $\mathbf{I}_{\text{det}} = [I_1,I_2,\ldots,I_{N_p}]$ where $N_p$ is the total number of pixels,
\begin{align} 
{I}_j = \int_{X_j} I_{\det}(x)dx, \label{eq:Ipixel} 
\end{align}
$X_j$ is the domain of the $j$th pixel i.e. $x \in [x_j - \Delta /2,x_j + \Delta/2)$, $\Delta$ is the pixel size and $x_j$  is the centre of the pixel. Accordingly the measured data vector takes the form $\mathbf{I}_{\text{meas}} = \mathbf{I}_{\text{det}} + \delta \mathbf{I}$ where $\delta\mathbf{I}$ is an unknown noise vector.

The expected precision of an observer's estimate of a parameter vector $\mathbf{w}$ can be conveniently parametrised using the covariance matrix, $\mathbb{K}_\mathbf{w}$, derived from, for example, repeated measurements. The Cram\'er-Rao lower bound (CRLB), however, states that the covariance matrix $\mathbb{K}_\mathbf{w}$ for an ideal observer is lower bounded by the inverse of the so-called Fisher information matrix (FIM) $\mathbb{J}_{\mathbf{w}}$ according to the matrix inequality \cite{Scharf1991} 
\begin{align}
\mathbb{K}_\mathbf{w} \geq \mathbb{J}_{\mathbf{w}}^{-1} ,\label{eq:CRLB}
\end{align}
where the FIM is defined as 
\begin{align}
\mathbb{J}_\mathbf{w} &= E\left[\left(\partialdiff{\ln f_\mathbf{I}(\mathbf{I}_{\text{meas}}|\mathbf{w})}{\mathbf{w}}\right)^T\partialdiff{\ln f_\mathbf{I}(\mathbf{I}_{\text{meas}}|\mathbf{w})}{\mathbf{w}}\right] \\
&=  E\left[\left(\partialdiff{}{\mathbf{w}} \left[\partialdiff{}{\mathbf{w}} \ln f_{\mathbf{I}}(\mathbf{I}_{\txtpow{meas}}|\mathbf{w})  \right]^T \right)^T\right] \label{eq:FIM_def},
\end{align}
$f_\mathbf{I}(\mathbf{I}_{\text{meas}}|\mathbf{w})$ is the probability density function (PDF) describing the conditional probability of measuring a particular value of $\mathbf{I}_{\text{meas}}$ for a given $\mathbf{w}$ and $E[\cdots]$ denotes the statistical expectation. The CRLB also implies the weaker set of inequalities $\sigma_{w_i}^2 \geq 1/[\mathbb{J}_\mathbf{w}]_{ii}$, where $\sigma_{w_i}^2$ is the estimation variance for each individual parameter $w_i$. It is important to note, however, that the CRLB as expressed by Eq.~\eqref{eq:CRLB} explicitly quantifies the uncertainty achievable by \emph{any} unbiased estimator and hence represents a fundamental limit to measurement precision, which can be asymptotically achieved using a maximum likelihood estimator \cite{Scharf1991}.

Using the chain rule, the FIM can be expressed in the alternative form \cite{Foreman2010}
\begin{align}
\mathbb{J}_\mathbf{w} = \left(\partialdiff{\mathbf{I}_{\text{det}}}{\mathbf{w}}\right)^T \mathbb{J}_{\mathbf{I}_{\text{det}}} \left(\partialdiff{\mathbf{I}_{\text{det}}}{\mathbf{w}}\right),\label{eq:FIM2}
\end{align}
where $\mathbb{J}_{\mathbf{I}_{\text{det}}}$ is defined analogously to \eqref{eq:FIM_def} in terms of the conditional PDF $f_\mathbf{I}(\mathbf{I}_{\text{meas}}|\mathbf{I}_{\text{det}})$. The appropriate choice of $f_\mathbf{I}(\mathbf{I}_{\text{meas}}|\mathbf{I}_{\text{det}})$ depends on the precise nature of noise present on a measurement. In inelastic optical spectroscopy the noise primarily derives from either the camera used to record the spectrum (e.g. due to dark current or read noise) or from the signal itself (in the form of shot noise). Shot noise is described by a Poisson PDF for which the noise variance scales with the mean intensity. Consequently, shot noise is more apparent at low signal levels and hence at shorter acquisition times. Given inelastic scattering is generally a weak process, shot noise is typically dominant. Although longer exposure times can be used to mitigate the effects of shot noise, this can come at the expense of increased dark noise which itself is also Poisson distributed. Gaussian distributed read out noise can dominate if  dark currents can then be sufficiently suppressed, for instance, by cooling the detector. In the case where read out noise is not the dominate noise source, however, a Gaussian PDF can still provide a good approximation to a Poisson PDF if the mean intensity is large enough. This shall be assumed to be the case for the majority of this work, albeit a short discussion of precision in inelastic spectroscopy in the presence of Poisson distributed noise is given below. At this point we thus make the simplifying assumption that the noise on each pixel is statistically independent and identically distributed according to a Gaussian PDF, i.e. 
\begin{align}
f_\mathbf{I}(\mathbf{I}_{\text{meas}}|\mathbf{I}_{\text{det}}) = \frac{1}{\sqrt{2\pi\sigma^2}}\exp\left[-\frac{\mathbf{I}_{\text{det}}\cdot \mathbf{I}_{\text{det}}}{2\sigma^2}\right]
\end{align}
from which it follows that $\mathbb{J}_{\mathbf{I}_{\text{det}}} = \mathbb{I}/\sigma^2$ where $\mathbb{I}$ is the $N_p \times N_p$ identity matrix.
Hence
\begin{align}
[\mathbb{J}_{\mathbf{w}}]_{kl} = \sum_{j}\! \frac{1}{\sigma^2} \partialdiff{I_j}{w_k}\partialdiff{I_j}{w_l}, \label{eq:FIMsum}
\end{align}
where $w_k$ is the $k$th element of $\mathbf{w}$ and $[\mathbb{J}_\mathbf{w}]_{kl}$ denotes the $(j,k)$th element of the FIM. The assumption of statistically independent and identically distributed noise implies that the dominant noise source is white in colour. Realistic noise sources, however, can frequently induce correlations between the noise for differing spectral components or on each pixel. Nevertheless, with knowledge of the power spectral density of the noise source a pre-whitening filter can be  constructed and applied, as is common in signal processing applications \cite{Scharf1991}. Moreover, coloured noise can be more fully treated within the framework of asymptotic Fisher information as has been discussed in Refs. \citenum{Zeira1990} and \citenum{Foreman2014}.

\section*{Numerical examples and limiting cases}\label{sec:examples}

Having established a theoretical framework to describe the obtainable precision in inelastic optical spectroscopy, we now consider a number of examples to illustrate the key experimental dependencies. Although the FIM can be evaluated numerically for the general case using Eqs.~\eqref{eq:Idet_convnew}, \eqref{eq:Ipixel} and  \eqref{eq:FIMsum}, we can also obtain analytic results for a number of limiting cases. Numerical examples will be restricted to consideration of the precision in determining $\bar{\Omega}$, as parametrised by $J_{\bar{\Omega},\bar{\Omega}} = J_{{\Omega},{\Omega}}/4$ only, since determination of the frequency shift $\Omega$ from inelastic scattering of light is the core task of inelastic optical spectroscopy, however, we will derive analytic results for all elements of the FIM.

\subsection*{Infinite-extent finely-pixelated detector : dispersion limited}\label{sec:CaseI}
We first consider the case in which pixelation of the detector is fine with respect to the spatial widths of any spectral features, i.e. $\Delta \ll \alpha \Gamma_p$, for $p = 0,\pm 1$. The detector is also assumed to be infinite in spatial extent.  The lineshape of the detected spectrum is assumed to be dictated by the Lorentzian lineshape of the underlying inelastic and Rayleigh peaks and the response function of the dispersive element.  Physically, this implies that the spatial width of the PSF, $\gamma_{\txtpow{opt}}$, is also much smaller than the spatial widths of any spectral features and that of the grating or VIPA, denoted $\gamma_{\txtpow{disp}}$, i.e. $\gamma_{\txtpow{opt}} \ll \alpha \Gamma_p$ and $\gamma_{\txtpow{opt}} \ll \gamma_{\txtpow{disp}}$. With these assumptions we can make the approximations $\bar{h}_{\txtpow{disp}}(x-x_0) = L(x';\alpha(\omega - \bar{\omega} + q\Omega_{\txtpow{FSR}})  + \bar{x} ,\gamma_{\txtpow{disp}})$,  $h_{\txtpow{opt}}(x) = \delta(x)$ whereby from Eq.~\eqref{eq:Idet_convnew}
\begin{align}
&I_{\txtpow{det}}^I(x) \approx N_I \sum_{p,q} |A_{p}B_q|^2
\int_0^\infty 
|L(\omega;\omega_0+p\Omega,\Gamma_p)|^2\left| L(x  ,  \bar{x}+  \alpha(\omega - \bar{\omega} + q \Omega_{\txtpow{FSR}} ), \gamma_{\txtpow{disp}}) \right|^2 d\omega.\label{eq:IdetI1}
\end{align}
Note that we use the sub- and superscript $I$ to distinguish this case and that we here use the normalisation constant $N_I$ (in distinction to $N$ used for the general case above) since we have applied an arbitrary scaling to $\bar{h}_{\txtpow{disp}}$ and $h_{\txtpow{opt}}(x)$ for mathematical convenience. Specifically, we have $N_I = I_\infty / \sum_{p=-1}^{+1} |A_p|^2$.
Noting $L(z-a;z_p,\zeta_p) = L(z,z_p + a,\zeta_p)$ and $L(z/a;z_p/a,\zeta_p/a) = \sqrt{a}L(z;z_p,\zeta_p)$ (as follow from inspection of Eq.~\eqref{eq:Lorentzian_om}), Eq.~\eqref{eq:IdetI1} can be transformed to
\begin{align}
&I_{\txtpow{det}}^I(x) \approx N_I \sum_{p,q} |A_{p}B_q|^2
\int_{-\infty}^\infty 
|L(\alpha\omega;\alpha(\omega_0+p\Omega),\alpha\Gamma_p)|^2\left| L(x -\alpha \omega ;  \bar{x}+  \alpha( q \Omega_{\txtpow{FSR}} - \bar{\omega} ), \gamma_{\txtpow{disp}}) \right|^2 d(\alpha\omega)\label{eq:IdetI2},
\end{align}
where we have also assumed that we are considering frequencies which are large compared to the relevant linewidths and spectral separations, such that we can safely extend the integration over $\omega$ to $-\infty$. Eq.~\eqref{eq:IdetI2} therefore shows that $I_{\txtpow{det}}^{I}(x)$ is given by the convolution of two Lorentzians, which is itself a Lorentzian function. Specifically we  find
\begin{align}
I_{\txtpow{det}}^I(x) &\approx N_I \sum_{p,q} |A_{p}B_q|^2
\left| L(x ;  \bar{x}+  \alpha\Omega_{pq},\alpha \Gamma_p +  \gamma_{\txtpow{disp}}) \right|^2 \label{eq:IdetI4}
\end{align}
where  $\Omega_{pq} = (p-\chi)\bar{\Omega}/2 + q \Omega_{\txtpow{FSR}}$. Now using Eq.~\eqref{eq:FIM2} we can express the elements of the FIM as
\begin{align}
[\mathbb{J}_{\mathbf{w}}]_{kl} \approx \!\frac{1}{\sigma^2 \Delta }\int_{-\infty}^{\infty} \!\!\partialdiff{I_j^I(x_j)}{w_k}\partialdiff{I_j^I(x_j)}{w_l} dx_j\label{eq:FIM3}
\end{align}
where the integration range on $x_j$ follows since we have assumed an infinite detector, 
\begin{align}
{I}_j^I &=  N_I\sum_{p,q} |A_{p}B_q|^2  \int_{X_j} \left|L\left(x ; \bar{x}  + \alpha\Omega_{pq}, \alpha \Gamma_p +  \gamma_{\txtpow{disp}}\right) \right|^2 dx\\
&\approx   N_I\sum_{p,q} |A_{p}B_q|^2  \Delta \left|L\left(x_j ; \bar{x}  + \alpha\Omega_{pq}, \alpha \Gamma_p +  \gamma_{\txtpow{disp}}\right) \right|^2 \label{eq:Ijapprox}
\end{align}
and the approximations hold in the fine pixelation limit.
Evaluating the derivatives in Eq.~\eqref{eq:FIM3} (remembering that $\bar{x} = \alpha (\bar{\omega} - \omega_{\txtpow{off}}$) where $\omega_{\txtpow{off}}$ is a pre-calibrated constant) we find
\begin{align}
&\partialdiff{I_j^I}{\bar{\omega}} \approx N_I\Delta\sum_{p,q}|A_{p}B_q|^2 \frac{\alpha }{\pi} \frac{  \gamma_p (x_j - \bar{x}  - \alpha\Omega_{pq}) }{[(x_j - \bar{x}  - \alpha\Omega_{pq})^2 + \gamma_p^2/4 ]^2 },\label{eq:diff_omegabar}\\
&\partialdiff{I_j^I}{\bar{\Omega}} \approx  N_I\Delta\sum_{p,q}|A_{p}B_q|^2\frac{\alpha}{2\pi}\frac{\gamma_p (p-\chi) (x_j - \bar{x}  - \alpha\Omega_{pq}) }{[ (x_j - \bar{x}  - \alpha\Omega_{pq})^2 +\gamma_p^2/4 ]^2 },\\
&\partialdiff{I_j^I}{\Gamma_{p}} \approx N_I\Delta\sum_{q } |A_{p}B_q|^2 \frac{\alpha}{2\pi}\frac{(x_j - \bar{x}  - \alpha\Omega_{pq})^2 - \gamma_p^2/4  }{[ (x_j - \bar{x}  - \alpha\Omega_{pq})^2 + \gamma_p^2/4 ]^2},\\
&\partialdiff{I_j^I}{|A_{p}|^2}  \approx  N_I\Delta \sum_{q } \frac{|B_q|^2}{\pi}\frac{\gamma_p/2}{ (x_j - \bar{x}  - \alpha\Omega_{pq})^2 + \gamma_p^2/4 },\label{eq:diff_Apq}
\end{align}
where $\gamma_p = \alpha \Gamma_p + \gamma_{\txtpow{disp}}$.
The form of the denominators in Eqs.~\eqref{eq:diff_omegabar}--\eqref{eq:diff_Apq} means that for a specific peak of order $(p,q)$ the derivative term is only non-negligible in the region of the peak. As such when evaluating the product of derivative terms in Eq.~\eqref{eq:FIM3} when $\Omega > \Gamma_0+\Gamma$ (i.e. the inelastic and Rayleigh peaks are well separated) it is reasonable to neglect any cross terms between peaks of different orders. With this approximation, substitution of Eqs.~\eqref{eq:diff_omegabar}--\eqref{eq:diff_Apq} into Eq.~\eqref{eq:FIM3}  allows the integration to be performed, yielding
\begin{align}
&J_{\bar{\omega},\bar{\omega}}^I = J_0\sum_{p=-1}^{+1} \frac{2 \alpha^2 |A_{p}|^4}{ \pi(\alpha \Gamma_p + \gamma_{\txtpow{disp}})^3} , \label{eq:JI1}\\
&J_{\bar{\Omega},\bar{\Omega}}^I = J_0\sum_{p=-1}^{+1} \frac{\alpha^2 (p-\chi)^2|A_{p}|^4}{2 \pi(\alpha \Gamma_p + \gamma_{\txtpow{disp}})^3} , \label{eq:JI2}\\
&J_{\Gamma_p,\Gamma_{p'}}^I = \delta_{pp'} \sum_{p'' = -1}^{+1} \delta_{|p||p''|}J_0\frac{\alpha^2 |A_{p''}|^4}{2\pi (\alpha \Gamma_{p''} + \gamma_{\txtpow{disp}})^3} ,\\
&J_{|A_{p}|^2,|A_{p'}|^2 }^I = \delta_{pp'}  \frac{J_0}{ \pi(\alpha \Gamma_p + \gamma_{\txtpow{disp}})} \label{eq:JIApAp}\\
&J_{\bar{\omega},\bar{\Omega}}^I =J_0 \sum_{p=-1}^{+1} \frac{(p-\chi)\alpha^2|A_{p}|^4}{\pi(\alpha \Gamma_p + \gamma_{\txtpow{disp}})^3} , \\
&J_{|A_{p}|^2,\Gamma_{p'} }^I = -\delta_{pp'} J_0\frac{\alpha|A_{p}|^2}{2 \pi(\alpha \Gamma_p + \gamma_{\txtpow{disp}})^2}  ,\label{eq:JI3}
\end{align}
where $J_0 = N_I^2 \Delta \sum_{q = -\infty}^\infty |B_q|^4 /  \sigma^2 $.
All other elements of the Fisher information matrix are zero, i.e. $J_{\bar{\omega},\Gamma_p}^I= J_{\bar{\Omega},\Gamma_p}^I = J_{\bar{\omega},|A_{p}|^2}^I = J_{\bar{\Omega},|A_{p}|^2}^I = 0$.

\begin{figure}[b!]
	\begin{center}
		\includegraphics[width=0.5\columnwidth]{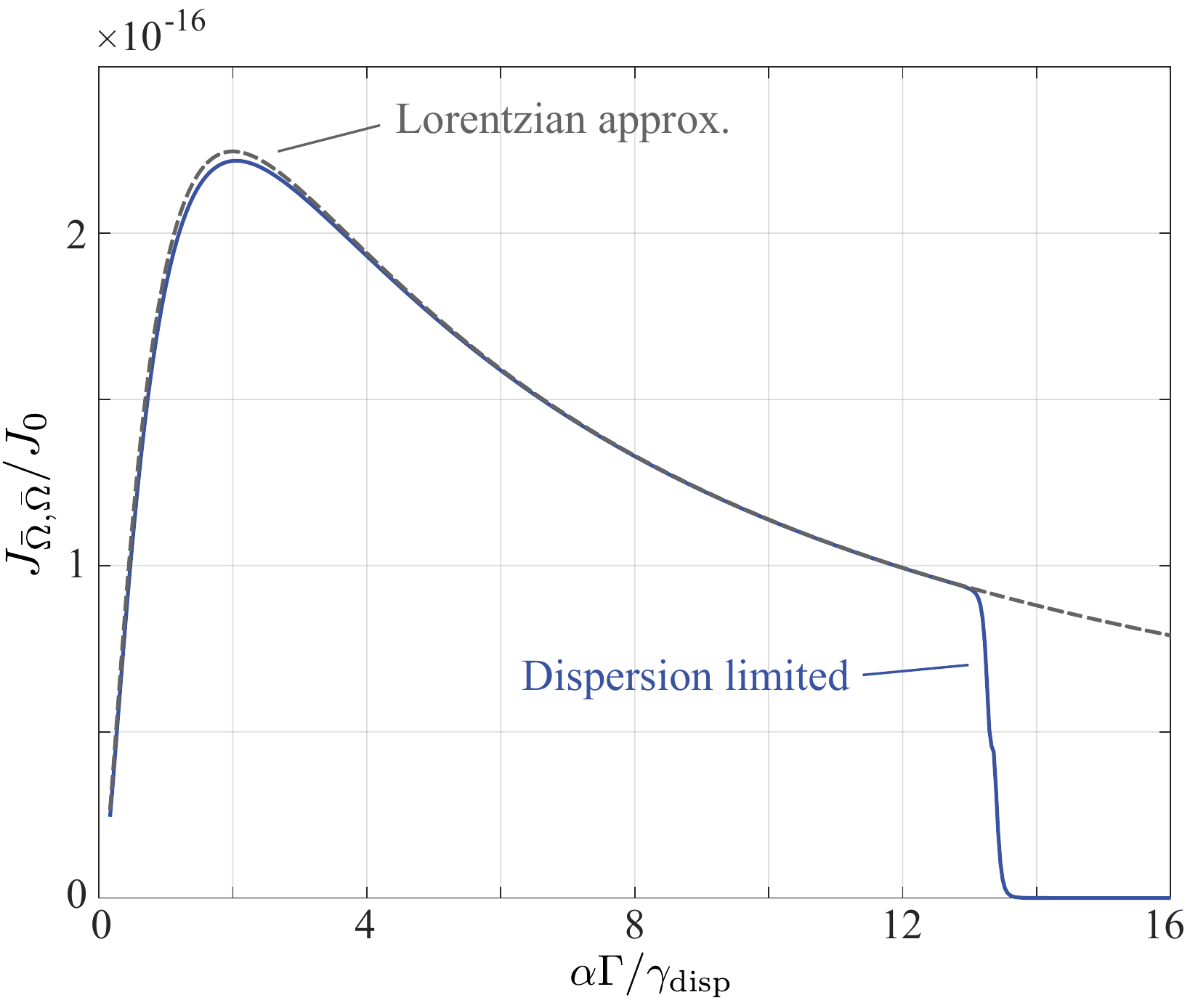}
		\caption{Numerical calculation (blue solid line) of  $J_{\bar{\Omega},\bar{\Omega}}$ and  the corresponding Lorentzian based approximation (dashed grey line). A finely pixelated detector with  $N_p = 5000$ pixels was assumed. The width of the response function of the dispersive element was taken as $\gamma_{\txtpow{disp}} = 8.4~\mu$m and the relay lens assumed to have a numerical aperture of $\text{NA} = 0.2$ such that the spectrometer is dispersion limited. See Table~\ref{tab:params} for other simulation parameters. \label{fig:Lorentz_approx}}
	\end{center}
\end{figure}  

For many applications, the inelastic frequency shift $\Omega$ (or equivalently $\bar{\Omega}$) is the primary parameter of interest. In Brillouin spectroscopy, for example, the Brillouin shift is proportional to the acoustic velocity of phonons in a material and can thus provide insights into mechanical properties of a sample \cite{Koski2005,Scarcelli2008}. The precision to which $\bar{\Omega}$ can be determined is parametrised by  $J_{\bar{\Omega},\bar{\Omega}}$. An example calculation of $J_{\bar{\Omega},\bar{\Omega}}$ is therefore shown in Figure~\ref{fig:Lorentz_approx}. Curves shown are for a full numerical calculation (solid blue) using Eqs.~\eqref{eq:Idet_convnew}, \eqref{eq:Ipixel} and  \eqref{eq:FIMsum} and for the approximate analytic result (dashed grey) given in Eq.~\eqref{eq:JI2}. For the numerical calculation the detector was necessarily finite in size and a relatively large numerical aperture of $\text{NA} = 0.2$ was used such that the PSF had a width of $\gamma_{\txtpow{airy}} = 1.4~\mu$m (or half a pixel), whereas the width of the dispersive element response function was set at 3 pixels (8.4~$\mu$m). Derivatives were calculated using a finite difference approximation. For simplicity only a single FSR was considered, i.e. $B_q = \delta_{q0}$ however it should be noted that this does not greatly affect our conclusions. All other simulation parameters are listed in Table~\ref{tab:params}, 
\begin{table}[t!]
	\begin{center}
		\begin{tabular}{l|c|c}
			Parameter & Symbol & Value \\\hline
			Size of CCD & $X$ & 14 mm\\ 
			Free spectral range & $\Omega_{\txtpow{FSR}}$ & 40~GHz\\
			Amplitude of Rayleigh peak & $A_0$ & 1 \\
			Amplitude of inelastic peaks & $A_{\pm 1}$  & 0.2 \\
			Spectral width of Rayleigh peak & $\Gamma_0$ & 0.2~GHz\\
			Spectral width of inelastic peaks & $\Gamma$ & 0.2~GHz\\
			Frequency shift & $\Omega$ & 12.5~GHz\\
			Rayleigh wavelength & $\lambda$ & 561~nm\\
			
		\end{tabular}
		\caption{Values of simulations parameters common to all examples. \label{tab:params}}
	\end{center}
\end{table}

The Fisher information in Figure~\ref{fig:Lorentz_approx} is plotted as a function of the ratio of the dimensionless parameter $\rho =\alpha \Gamma/\gamma_{\txtpow{disp}}$ which describes the ratio of the intrinsic spatial width relative to the width of the dispersive element's response function. 
Good agreement between the numerical and approximate results is generally evident, except that the Fisher information (FI) calculated numerically drops to zero at $\alpha \Gamma/\gamma_{\txtpow{disp}}\sim 13$. This discrepancy will be discussed further below and arises because our numerical calculations necessarily consider a finite sized detector in contrast to the assumption made in our theoretical analysis.

Physical insight into the behaviour shown in Figure~\ref{fig:Lorentz_approx} and indeed into Eqs.~\eqref{eq:JI1}--\eqref{eq:JI3} can be gained by first considering the case of  a purely monochromatic input, but allowing for the finite width of the amplitude response function of the dispersive element ($\Gamma_p = 0$, $\gamma_{\txtpow{disp}}\neq 0$). In this case the contribution to Eqs.~\eqref{eq:JI1}--\eqref{eq:JI3} from each spectral peak follows an inverse power law (of varying degree) in the instrumental peak width $\gamma_{\txtpow{disp}}$.  As the spatial width of a peak decreases, so the energy contained within that peak is confined to a smaller area, such that the signal to noise ratio at each position on the detector improves and a better estimation precision ultimately results.  In each non-zero element of the FIM, there is however an additional dependence on the scale factor $\alpha$ appearing in the numerator (with the exception Eq.~\eqref{eq:JIApAp}). This factor captures the intuitive expectation that spatial positions, separations or widths can be more precisely determined when they are magnified. Taking the estimation of $\bar{\Omega}$ as an illustrative example, we note that as $\alpha$ decreases so the inelastic peaks are positioned closer together on the detector, however, the finite (fixed) peak width means it is consequently harder to individually resolve the peaks and hence determine their separation. 

Similarly, when considering an ideal spectrometer, but allowing for a finite intrinsic spectral width ($\gamma_{\txtpow{disp}} = 0, \Gamma_p \neq 0$), we find that the non zero elements of the FI decrease as the spatial width, $\alpha \Gamma_p$, of the observed peaks increases. In this case however, since the intrinsic spatial width decreases with $\alpha$, the difficulty in resolving individual peaks is somewhat mitigated. In the general dispersion limited case ($\gamma_{\txtpow{disp}} \neq 0, \Gamma_p \neq 0$) the total observed linewidth is dictated by both the intrinsic spectral width and the broadening caused by the dispersive element, as reflected by the aggregate width $\alpha \Gamma_p + \gamma_{\txtpow{disp}}$ appearing in the denominators of Eqs.~\eqref{eq:JI1}--\eqref{eq:JI3}, however the principles dictating the estimation precision are the same. Finally we note that whilst Eqs.~\eqref{eq:JI1}--\eqref{eq:JI3} predict that  the obtainable FI tends to infinity as $\alpha \Gamma_p + \gamma_{\txtpow{disp}} \rightarrow 0$, this information divergence is of no physical relevance since the widths of the peaks on the detector become comparable to the pixel size in this limit hence invalidating our assumption of a finely pixelated detector. This case is considered in greater detail below.

The relative magnitude of $\alpha$ and $\gamma_{\txtpow{disp}}$ can  be experimentally controlled, for example, by varying the magnification factor of the relay lens, so as to achieve an optimal balance between the two competing effects hence yielding a maximum FI (or equivalent the best obtainable precision).  When considering inelastic peaks of equal amplitude $A$ Eq.~\eqref{eq:JI2}, for example, can be written in the form
\begin{align}
J_{\bar{\Omega},\bar{\Omega}}^I =\frac{ J_0|A|^4}{ \pi\gamma_{\txtpow{disp}} \Gamma^2 } \frac{\rho^2}{(1+\rho)^3}
\end{align}
such that a maximum FI of $4 J_0 |A|^4 / (27 \pi \gamma_{\txtpow{disp}} \Gamma^2)  $ can be obtained when $\rho=2$, or equivalently when the intrinsic spatial inelastic peak width is twice that of the amplitude response function of the dispersive element $\alpha\Gamma = 2 \gamma_{\txtpow{disp}}$.
Similar maxima are also found in $J_{\bar{\omega},\bar{\omega}}$ and $J_{\Gamma_p,\Gamma_{p'}}$, whereas the maximum correlation between estimates of $|A_p|^2$ and $\Gamma_{p}$ (as described by $J_{|A_{p}|^2,\Gamma_{p'} }$) occurs when $\rho = 1$ or equivalently $\alpha\Gamma = \gamma_{\txtpow{disp}} $. There exists no optimal configuration for the remaining elements of the FIM.

\subsection*{Infinite-extent finely-pixelated detector : diffraction limited}
In a similar vein to above we can determine the FIM for an infinite-extent, finely pixelated detector, however, instead of assuming the intensity distribution on the detector is limited by the response function of the dispersive element, we can instead consider the case in which the PSF, $h_{\txtpow{opt}}(x)$, of the relaying optics  dominates. For a one-dimensional case without aberrations, the finite numerical aperture $\text{NA}$ of the optics implies that $h_{\txtpow{opt}}(x) \sim \text{sinc}(\pi x /\gamma_{\txtpow{Abbe}})$ where $\gamma_{\txtpow{Abbe}} = \lambda / (2\text{NA})$ determines the position of the first zero of $h(x)$. For simplicity, however, we approximate the instrument response function by a Gaussian distribution, 
\begin{align}
G(x;x_p,\gamma_{\txtpow{opt}})&= \frac{1}{(2\pi \gamma_{\txtpow{opt}}^2)^{1/4}} \exp\left[-\frac{(x-x_p)^2}{4\gamma_{\txtpow{opt}}^2}\right],\label{eq:Gaussian_om}
\end{align}
with a spatial width of $\gamma_{\txtpow{opt}}$ chosen so as to match the full-width half-maximum of the sinc function, implying  $\gamma_{\txtpow{opt}}= 1.89549 \gamma_{\txtpow{Abbe}} / (2\pi \sqrt{\ln 2}) = 0.36235 \gamma_{\txtpow{Abbe}} $. For this case we thus make the approximations: $\bar{h}_{\txtpow{disp}}(x'-x_0) = \delta(x' - \alpha(\omega - \bar{\omega} + q \Omega_{\txtpow{FSR}})  - \bar{x} )$ and  $h_{\txtpow{opt}}(x) = G(x;0,\gamma_{\txtpow{opt}})$, whereby
\begin{align}
&I_{\txtpow{det}}^{II}(x) 
=N_{II}\sum_{p,q} |A_{p}B_{q}|^2
\int_0^\infty |L(\omega;\omega_0+p\Omega,\Gamma_p)|^2\left|  G(\alpha (\omega - \bar{\omega} + q \Omega_{\txtpow{FSR}})  +\bar{x} -x;0,\gamma_{\txtpow{opt}}) \right|^2 d\omega .
\end{align}
Again we use the normalisation constant $N_{II}$  since we have scaled $\bar{h}_{\txtpow{disp}}$ and $h_{\txtpow{opt}}(x)$ for convenience where we find $N_{II} = N_I $.

Once more using the properties of Lorentzians as above in addition to  similar properties for Gaussian lineshapes  we can write the resulting intensity distribution as the convolution of a Lorentzian and a Gaussian lineshape, i.e. 
\begin{align}	
\!\!I_{\txtpow{det}}^{II}(x)&=N_{II}\sum_{p,q} |A_{p}B_{q}|^2 V(x;\bar{x} + \alpha \Omega_{pq},\alpha\Gamma_p,\gamma_{\txtpow{opt}})
\end{align}
where $V(x;\bar{x} + \alpha \Omega_{pq},\alpha\Gamma_p,\gamma_{\txtpow{opt}})$ is the Voigt profile \cite{Ida2000}. For simplicity we do not consider the full integral form for the Voigt profile, but instead restrict attention to the pseudo-Voigt distribution described in \cite{Ida2000} whereby
\begin{align}
&V(x;\bar{x} + \alpha \Omega_{pq},\alpha\Gamma_p,\gamma) \approx \eta_p |L(x;\bar{x} + \alpha \Omega_{pq}, \beta_p/\alpha )|^2  + (1-\eta_p)|G(x;\bar{x} + \alpha \Omega_{pq},\beta_p / (2\sqrt{2\ln 2}))|^2 \label{eq:pseudoVoigt}
\end{align}
where $\mu_p = {\beta_{L,p}}/{\beta_p}$, 
\begin{align}
\eta_p &= 1.36603 \mu_p - 0.47719 \mu_p^2 + 0.11116 \mu_p^3,\\
\beta_p &= \left[\beta_G^5 + 2.69269 \beta_G^4 \beta_{L,p} +  2.42843 \beta_G^3 \beta_{L,p}^2 \quad + 4.47163 \beta_G^2 \beta_{L,p}^3 + 0.07842 \beta_G \beta_{L,p}^4 + \beta_{L,p}^5\right]^{1/5}
\end{align}
and the FWHM of the Lorentzian and Gaussian intensity distributions are $\beta_{L,p}=2\alpha\Gamma_p$ and $\beta_G=2\sqrt{2\ln 2} \gamma_{\txtpow{opt}}$ respectively.
Using this approximation we rewrite $I_{j} = \sum_{p,q} \eta_p I_{pq}^L(x_j) + (1-\eta_p) I_{pq}^G(x_j)$ where $I_{pq}^L(x_j)$ and $I_{pq}^G(x_j)$ derive from the Lorentzian and Gaussian terms of the $(p,q)$th order peak. Again neglecting any cross term between adjacent peaks and also neglecting any parameter dependence of $\eta$ we have that
\begin{align}
\partialdiff{I_j^{II}}{w_k}\partialdiff{I_j^{II}}{w_l} &\approx \eta_p^2 \partialdiff{I_{pq}^L}{w_k}\partialdiff{I_{pq}^L}{w_l}  + (1-\eta_p)^2\partialdiff{I_{pq}^G}{w_k}\partialdiff{I_{pq}^G}{w_l} + \eta_p(1-\eta_p)\left[\partialdiff{I_{pq}^G}{w_k}\partialdiff{I_{pq}^L}{w_l} + \partialdiff{I_{pq}^L}{w_k}\partialdiff{I_{pq}^G}{w_l} \right],
\end{align} 
where the dependence of $I_{pq}^{\cdots}$ on $x_j$ has been suppressed for clarity.
It then follows that the FIM can be partitioned into three contributions viz. $\mathbb{J}^{{II}}_{\mathbf{w}}  =  \mathbb{J}^L_{\mathbf{w}} +  \mathbb{J}^G_{\mathbf{w}} + \mathbb{J}^{GL}_{\mathbf{w}}$. The first term, $\mathbb{J}_\mathbf{w}^L$, will take the same form as Eqs.~\eqref{eq:JI1}--\eqref{eq:JI3} with the replacement $\gamma_{\txtpow{disp}}\rightarrow 0$ and with an additional factor of $\eta_p^2$ within the summations. The FIM associated with the second term can be evaluated by following the same logic as in the previous section ultimately yielding
\begin{align}
&J_{\bar{\omega},\bar{\omega}}^G = J_0\sum_{p=-1}^{+1} (1-\eta_p)^2\frac{\alpha^2|A_{p}|^4 }{4\sqrt{\pi}\gamma_{\txtpow{opt}}^3} ,\label{eq:JG1}\\
&J_{\bar{\Omega},\bar{\Omega}}^G =J_0 \sum_{p=-1}^{+1} (1-\eta_p)^2\frac{ \alpha^2 (p-\chi)^2|A_{p}|^4}{16\sqrt{\pi}\gamma_{\txtpow{opt}}^3}\label{eq:JG2},\\
&J_{|A_{p}|^2,|A_{p'}|^2 }^G = \delta_{pp'}J_0  (1-\eta_p)^2\frac{ \alpha^2}{2\sqrt{\pi} \gamma_{\txtpow{opt}}},\\
&J_{\bar{\omega},\bar{\Omega}}^G =J_0\sum_{p=-1}^{+1}(1-\eta_p)^2 \frac{ \alpha^2(p-\chi)|A_{p}|^4}{8\sqrt{\pi}\gamma_{\txtpow{opt}}^3}
\end{align}
and $J_{\Gamma_p,\Gamma_{p'}}^G  = J_{|A_{p}|^2,\Gamma_{p'}}^G = J_{\bar{\omega},\Gamma_p}^G= J_{\bar{\Omega},\Gamma_p}^G = J_{\bar{\omega},|A_{p}|^2}^G = J_{\bar{\Omega},|A_{p}|^2}^G = 0$.
The cross FIM terms follow similarly and are given by
\begin{align}
&{J}^{GL}_{\bar{\omega},\bar{\omega}} = J_0 \sum_{p=-1}^{+1}\eta_p(1-\eta_p)|A_{p}|^4  \frac{\sqrt{2}\alpha^2}{\pi  \gamma_{\txtpow{opt}}^3} f(r_p),\\
&{J}^{GL}_{\bar{\Omega},\bar{\Omega}} = J_0 \sum_{p=-1}^{+1}\eta_p(1-\eta_p)|A_{p}|^4 \frac{\alpha^2 (p-\chi)^2}{2\sqrt{2}\pi  \gamma_{\txtpow{opt}}^3}  f(r_p), \label{eq:JGL2}\\
&{J}^{GL}_{\bar{\omega},\bar{\Omega}} =J_0 \sum_{p=-1}^{+1}\eta_p(1-\eta_p)|A_{p}|^4   \frac{(p-\chi)\alpha^2}{\sqrt{2}\pi  \gamma_{\txtpow{opt}}^3}  f(r_p),\\
&{J}^{GL}_{|A_{p}|^2,\Gamma_{p'} }  = \delta_{pp'}  J_0\eta_p(1-\eta_p)|A_{p}|^2 \frac{\alpha}{2\pi \gamma_{\txtpow{opt}}^2} g(r_p),\\
&{J}^{GL}_{|A_{p}|^2,|A_{p'}|^2 } = \delta_{pp'}J_0\eta_p(1-\eta_p)\sqrt{\frac{2}{\pi}}\frac{w\left(r_p\right)  }{\gamma_{\txtpow{opt}}} \label{eq:JGL5},
\end{align}
where $r_p = {\alpha \Gamma_p}/({2\sqrt{2}\gamma_{\txtpow{opt}}})$,
\begin{align}
f(z) &= \sqrt{\pi}\left(2z^2 + 1\right) w\left(z\right) -2z ,\\
g(z) &= \sqrt{\pi} z w\left(z\right)  - 1,\\
w(z) &= \exp[z^2]\text{erfc}[z]
\end{align}
and all other elements of $\mathbb{J}^{GL}_{\mathbf{w}} $ are zero.

\begin{figure}[t!]
	\begin{center}
		\includegraphics[width=0.5\columnwidth]{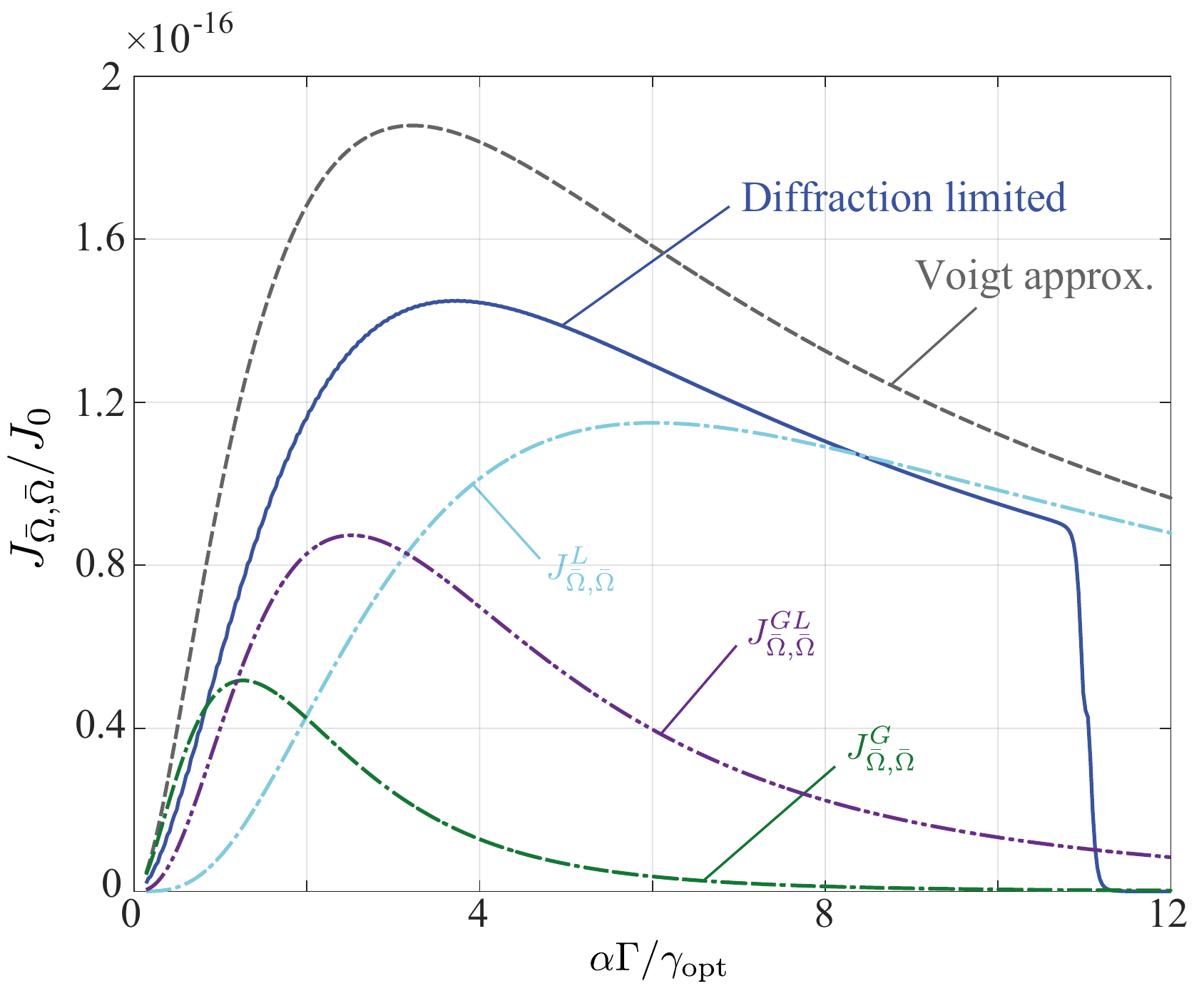}
		\caption{Numerical calculation (blue solid line) of  $J_{\bar{\Omega},\bar{\Omega}}$ and  the corresponding Voigt based approximation (dashed grey line). A finely pixelated detector with  $N_p = 5000$ pixels was assumed. The width of the response function of the dispersive element was taken as $\gamma_{\txtpow{disp}} = 8.4~$nm and the relay lens assumed to have a numerical aperture of $\text{NA} = 0.01$ such that the spectrometer is diffraction limited. See Table~\ref{tab:params} for other simulation parameters.  Individual contributions to the Voigt approximation are shown by the dot-dashed light blue, green and purple lines  corresponding to Eqs.~\eqref{eq:JI2} (see also text), \eqref{eq:JG2} and \eqref{eq:JGL2}  respectively. \label{fig:Voigt_approx}}
	\end{center}
\end{figure}  

\begin{figure}[t!]
	\begin{center}
		\includegraphics[width=0.5\columnwidth]{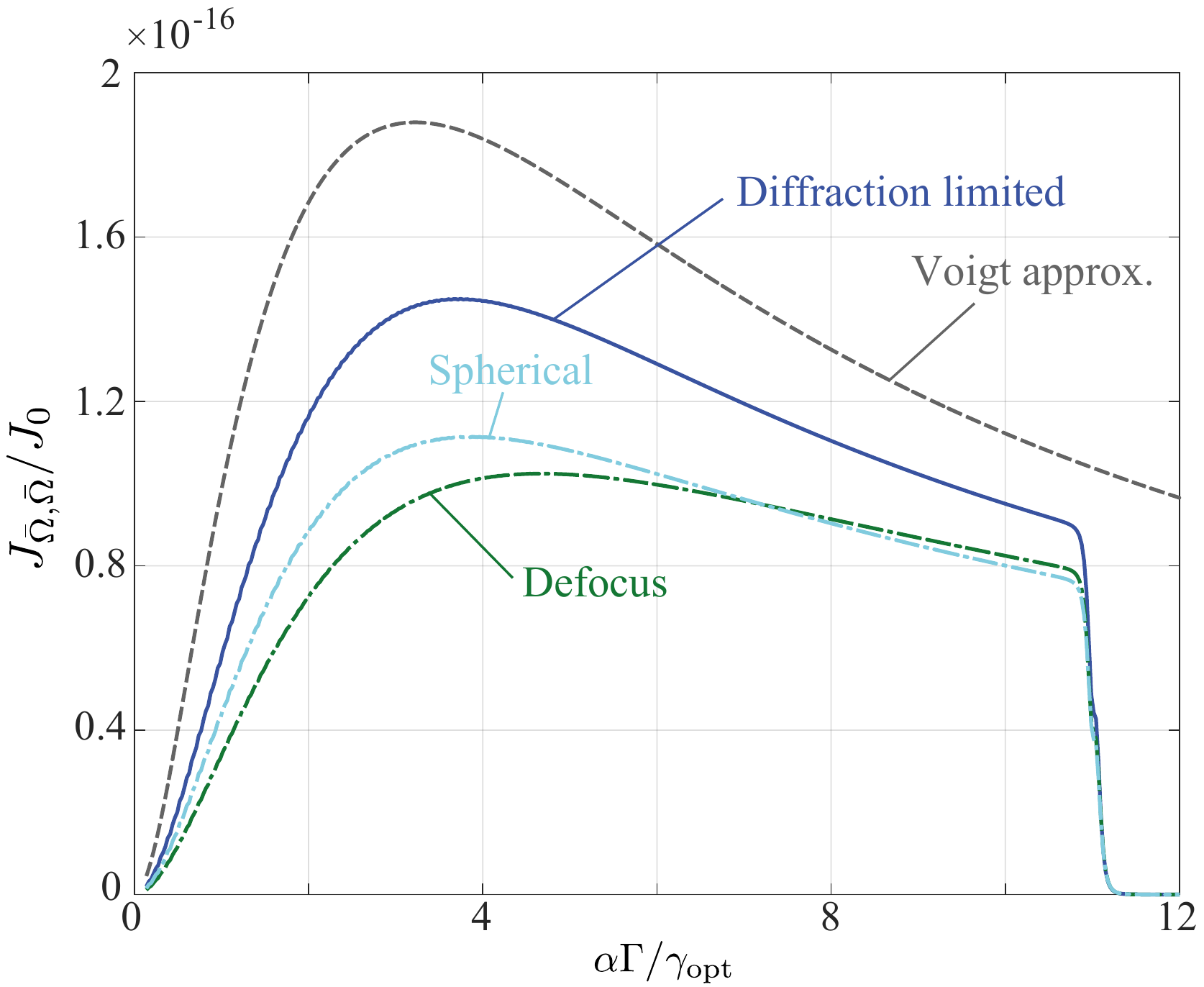}
		\caption{Numerical results (blue solid  curve) for the diffraction limited case (as per. Figure~\ref{fig:Voigt_approx}), with the addition of one wave of defocus (green dash-dotted curve) and spherical aberration (light blue dash-dotted curve). The Voigt based approximation is also shown by the grey dashed curve. \label{fig:aberrations}}
	\end{center}
\end{figure}

Numerical results comparing the calculated FI $J_{\bar{\Omega},\bar{\Omega}}$ to the analytic Voigt based approximation for the diffraction limited case are shown in Figure~\ref{fig:Voigt_approx}, now plotted as a function of the ratio $\tau = \alpha \Gamma/\gamma_{\txtpow{opt}}$. Simulation parameters are again given in Table~\ref{tab:params}. Additionally a numerical aperture of $\mbox{NA} = 0.01$ corresponding to $\gamma_{\txtpow{opt}} \approx 3.6$ pixels was used, whilst a negligible value of $\gamma_{\txtpow{disp}} = 8.4$~nm was assumed. Individual contributions to the analytic result from $\mathbb{J}^L$, $\mathbb{J}^G$ and $\mathbb{J}^{GL}$ are also shown. Good qualitative agreement between the numerical and approximate results are seen, however, for small $\tau$ numerical discrepancies are relatively large. This discrepancy is a result of the Gaussian approximation used to represent the PSF, with the exact functional form playing a more critical result in this regime (as reflected by the relative importance of $\mathbb{J}^G$ and $\mathbb{J}^{GL}$). Aberrations present in the relay optics which alter the shape of the PSF (i.e. excluding tilt and piston)  would thus be expected to have a significant effect at small $\tau$ (e.g. from use of larger magnifications), as is indeed borne out in calculations as shown in Figure~\ref{fig:aberrations}. Specifically, we have plotted the variation of $J_{\bar{\Omega},\bar{\Omega}}$ using the same  parameters used for Figure~\ref{fig:Voigt_approx}, however, with the addition of one wave of defocus or spherical aberration. For each case the reduction in the obtainable FI, resulting from the overall PSF broadening, is similar in each case, however, the shift in the optimal $\tau$ is relatively small.

\begin{figure}[t!]
	\begin{center}
		\includegraphics[width=0.45\columnwidth]{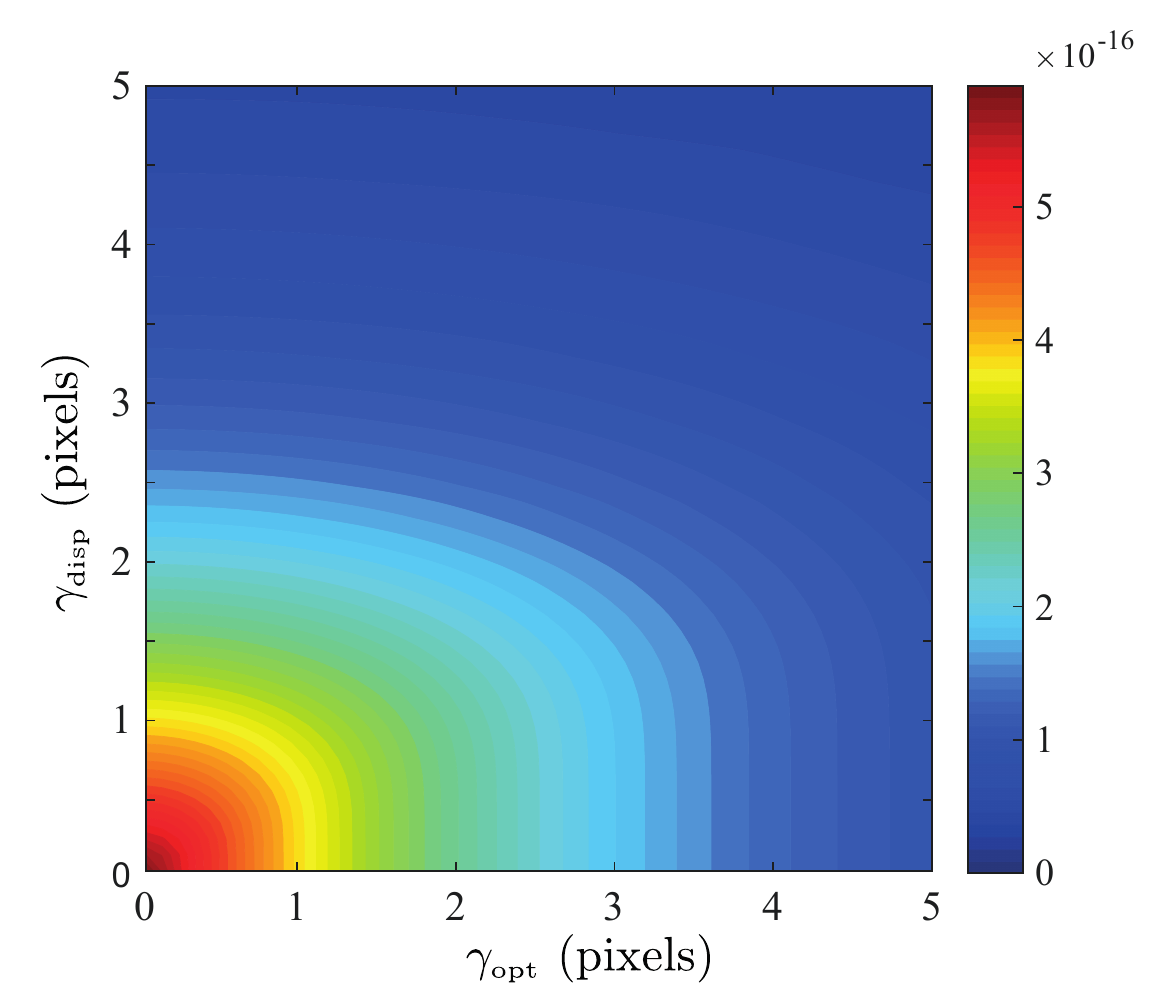}
		\caption{Variation of $J_{\bar{\Omega},\bar{\Omega}}$ as a function of the width of the PSF and dispersive element response function $\gamma_{\txtpow{opt}}$ and $\gamma_{\txtpow{disp}}$ assuming $\alpha \Omega_{\txtpow{FSR}} / X= 0.25$.  \label{fig:widths}}
	\end{center}
\end{figure}

Figure~\ref{fig:widths} shows the numerical variation of $J_{\bar{\Omega},\bar{\Omega}}$ for arbitrary values of $\gamma_{\txtpow{disp}}$ and $\gamma_{\txtpow{opt}}$ in the aberration free case, whereby it is seen that the obtainable precision monotonically decreases for fixed $\alpha$ as either response function broadens.

\subsection*{Finite-extent finely-pixelated detector\label{sec:finitedetector}}
The cases considered hitherto have assumed an infinite detector. Accordingly, when determining the elements of the FIM, the summation over each data point $I_j$ (defined by Eq.~\eqref{eq:Ipixel}) could be accurately modelled by an integration over an infinite domain, as was done in Eq.~\eqref{eq:FIM3}. Upon considering the more realistic case of a detector of finite spatial extent $X$ (albeit still finely pixelated), the integration domain in Eq.~\eqref{eq:FIM3} must be restricted to $X$. Assuming that any given spectral feature does not straddle the edge of the detector, the effect of the finite integration domain is to limit the summation over $q$ (and potentially $p$). Specifically, denoting the spatial position of the spectral peak indexed by $p$ and $q$ as $x_{pq} = \bar{x} + \alpha \Omega_{pq}$ and its associated experimental width by $\gamma$, the summations appearing in Eqs.~\eqref{eq:JI1}--\eqref{eq:JI3} and \eqref{eq:JG1}--\eqref{eq:JGL5} are only over peaks for which $\text{max}[|x_{pq} \pm \gamma|] \lesssim X/2 $. No information is obtained from peaks falling beyond the spatial extent of the detector as would be expected by intuition thus accounting for the drop in the calculated FI to zero seen in Figures~\ref{fig:Lorentz_approx}--\ref{fig:aberrations}. Only partial information is obtained for peaks which straddle the edge of the detector. Moreover, only a single FSR was considered in our discussion thus far, however, when multiple FSRs are present, a staggered fall off in the FI to zero is seen, with each step occurring when a single peak moves out of the detection area.

\subsection*{Coarse pixelation}\label{sec:pixelation}
To illustrate the effect of pixelation on the obtainable estimation accuracy it is sufficient to consider the FI obtained from measurement of a single peak of the intensity distribution falling on the detector. We assume that the detector has $N_p$ pixels indexed by $j = 1,2,\ldots, N_p$. Due to our assumption that each peak does not overlap significantly, the total FI then follows by summing the information obtained for each individual peak (c.f. for example Eqs.~\eqref{eq:JI1}--\eqref{eq:JI3}). For simplicity we shall assume a Lorentzian lineshape of width $\gamma$, such that 
\begin{align}
I_{\txtpow{det}}(x) = |A_p|^2| L(x;\bar{x} + \alpha(p-\chi)\Omega,\gamma)|^2.
\end{align}
In the extreme case of coarse pixelation we can assume that the pixels are so large that a single spectral peak spans only three pixels before falling to negligible intensities. The measured data values $I_j$ are thus zero unless $n-1 \leq j \leq n + 1$, where $n$ is the index of the centre pixel of the three under consideration. Since we have assumed that $ I_{\txtpow{det}}(x) \approx 0$ on all but three pixels we can extend the integration domains appearing in Eq.~\eqref{eq:Ipixel}, such that 
\begin{align}
I_j &= \delta_{j,n-1}\int_{-\infty}^{x_{n } - \Delta/2} I_{\txtpow{det}}(x)dx  +  \delta_{j,n}\int_{x_{n } - \Delta/2}^{x_{n } + \Delta/2} I_{\txtpow{det}}(x)dx +  \delta_{j,n+1}\int_{x_{n } + \Delta/2}^{\infty} I_{\txtpow{det}}(x) dx.
\end{align}
For the Lorentzian lineshape assumed, the integration can be performed analytically yielding
\begin{align}
I_{n\pm 1} &= \frac{|A_p|^2}{\pi} \left[\frac{\pi}{2} \pm \text{arctan}\left( \frac{x_n - x_p \mp \Delta/2}{\gamma / 2}\right)\right] 
\end{align}
and $I_n = |A_p|^2 - I_{n-1} - I_{n+1}$,  where $x_p =  \alpha (p-\chi) \Omega + \bar{x}$.
Letting $I_{n} \triangleq |A_p|^2 i_{p,n}$, for a single FSR we have 
\begin{align}\textsl{}
I_j &= |A_{-1}|^2(i_{-1,l-1} \delta_{j,l-1} + i_{-1,l}\delta_{j,l} + i_{-1,l+1}\delta_{j,l+1} )  + |A_0|^2(i_{0,m-1} \delta_{j,m-1} + i_{0,m}\delta_{j,m} + i_{0,m+1}\delta_{j,m+1} ) \nonumber \\
&\quad\quad+ |A_{+1}|^2(i_{+1,n-1} \delta_{j,n-1} + i_{+1,n}\delta_{j,n} + i_{+1,n+1}\delta_{j,n+1} )  \nonumber
\end{align}
where $l,m,n$ denote the indices of the central pixel for the $p=-1,0,1$ order peak respectively. An element of the FIM then follows as
\begin{align}
[\mathbb{J}]_{i,k} = \frac{1}{\sigma^2} \partialdiff{^2}{w_i\partial w_k}\sum_{p=-1}^1 |A_p|^4 i_{p,t(p)}^2 \label{eq:Jpixel1}
\end{align}
where we have  used the equivalent definition of FIM given in Eq.~\eqref{eq:FIM_def}. It is important to note that in Eq.~\eqref{eq:Jpixel1} we have introduced the mapping function $t(p)$. In particular $t(p)$ is the integer satisfying the inequality
\begin{align}
x_{t} - \Delta/2 < \alpha (p-\chi) \Omega + \bar{x} \leq x_t + \Delta/2 .
\end{align}
To gain further insight we temporally redefine the pixel index $j$ such that the $j$th pixel is centred at $x=0$, whereby $t = -\text{floor}[X/(2\Delta)],\ldots,-1,0,1,\ldots \text{ceil}[X/(2\Delta)]$. Hence $x_t = t \Delta$ and $t$ is the integer satisfying $t\Delta - \Delta/2 < x_p \leq t\Delta + \Delta/2 $. Thus ${x_p}/{\Delta} - {1}/{2} \leq t < {x_p}/{\Delta} + {1}/{2}$ implying 
\begin{align}
t = \text{round}\left[\frac{x_p}{\Delta} - \frac{1}{2}\right] = \text{round}\left[\frac{\alpha (p-\chi) \Omega}{\Delta} + \frac{\bar{x}}{\Delta} - \frac{1}{2}\right] .
\end{align}
\begin{figure}[t!]
	\begin{center}
		\includegraphics[width=0.5\columnwidth]{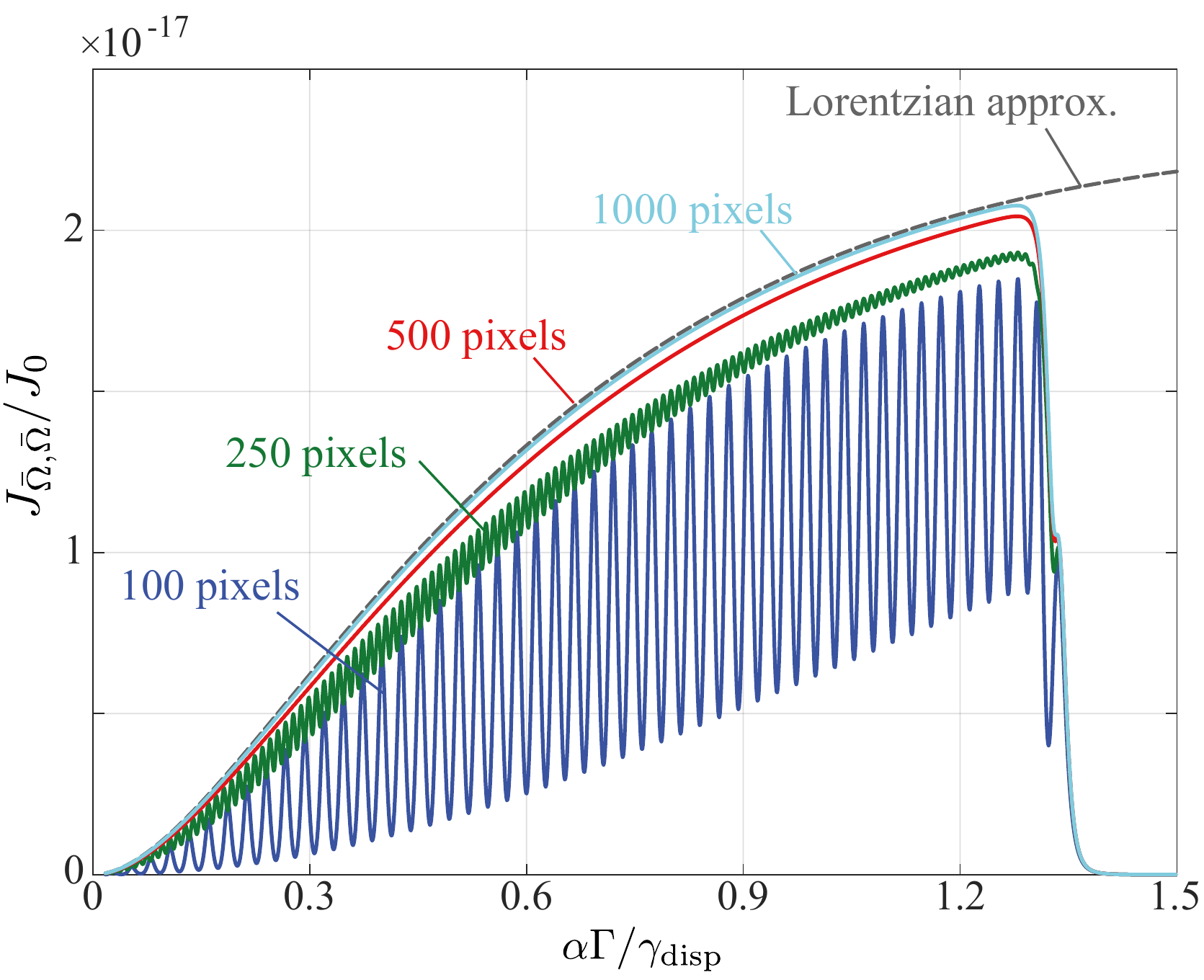}
		\caption{Calculated Fisher information in the coarse pixelation limit. Peak widths of $\gamma_{\txtpow{disp}} = 84~\mu$m  and $\gamma_{\txtpow{opt}} = 8.4~\mu$m were assumed, whilst the number of pixels on a finite size detector was varied. Other simulation parameters are given in Table~\ref{tab:params}.   \label{fig:pixels}}
	\end{center}
\end{figure}  
Typically $\Omega/\Delta \gg 1$ implying that $t$, and hence the acquired FIM, oscillates rapidly as a function of $\alpha$. Since $\alpha$ describes the linear mapping between the frequency and spatial domain, this means that the obtainable FIM, and hence the obtainable precision, is strongly dependent on the angular dispersion of the spectrometer and the optical magnification in the coarse pixelation regime. This behaviour is to be expected because in this regime the details of the spectral peak are barely resolvable by the detector. Similarly, the obtainable precision depends strongly on the registration of the incident spectrum with respect to the detector pixels, as parametrised by $\bar{x}$. To illustrate the oscillations we have numerically calculated $J_{\bar{\Omega},\bar{\Omega}}$ for a detector with varying numbers of pixels. Specifically we use the parameters given in Table~\ref{tab:params} and consider a detector with 100, 250, 500 and 1000 pixels (corresponding to pixel sizes of 140, 56, 28, 14~$\mu$m respectively).  The width of the dispersive element's response function was fixed at $\gamma_{\txtpow{disp}} = 84~\mu$m, whilst the numerical aperture of the relay lens was set such that $ \gamma_{\txtpow{opt}} = \gamma_{\txtpow{disp}}/10$. With these parameters the simulated spectrometer operates within the dispersion limited regime. Numerical results are shown in Figure~\ref{fig:pixels}. For the largest pixel size (corresponding to $\gamma_{\txtpow{disp}} = 0.6$ pixels) large oscillations  in $J_{\bar{\Omega},\bar{\Omega}}$  are observed. The magnitude of these oscillations decrease as the pixel size decreases until the finely pixelated regime is reached. We note that in these simulations the larger choice of $\gamma_{\txtpow{disp}}$ means that the optimal configuration ($\rho = 2$) requires a scaling factor $\alpha$ for which the inelastic spectral peaks lie beyond the detector and no information regarding $\Omega$ can be obtained. In turn this means that the peak in  $J_{\bar{\Omega},\bar{\Omega}}$ is not seen (as compared to e.g. Figure~\ref{fig:Lorentz_approx}). Instead the optimal $\alpha$ is that for which the inelastic peaks  lie just within the spatial extent of the detector (assuming that the Rayleigh peak is centred).

Within any experimental context it is desirable to avoid oscillations in the obtainable precision and thus it is useful to estimate the pixel size at which coarse pixelation effects become relevant. To do so we consider the intensity recorded on a single pixel $j$ for a single Lorentzian peak. For a peak of arbitrary width $\gamma$ we have
\begin{align}
I_j &= \int_{x_j + \Delta/2}^{x_{j } - \Delta/2} I_{\txtpow{det}}(x)dx  =  \frac{|A_p|^2}{\pi} \left[\text{arctan}\left( \frac{x_j - x_p + \Delta/2}{\gamma / 2}\right) - \text{arctan}\left( \frac{x_j - x_p - \Delta/2}{\gamma / 2}\right)\right] 
\end{align}
where the integration has been performed analytically in a similar fashion to above. Performing a Maclaurin expansion in terms of the ratio of the pixel size and linewidth, i.e. $\Delta /\gamma $, yields
\begin{align}
I_j \approx \Delta \, I_{\txtpow{det}}(x_j) + O[(\Delta/\gamma)^3].
\end{align}
To lowest order in $\Delta/\gamma$ we can thus consider each pixel reading to be a discrete sample of the underlying lineshape (as per intuitive expectations), whereby the problem of coarse pixelation can be analysed in terms of under sampling of the incident intensity distribution. In particular, we first note that  the power spectrum of a Lorentzian lineshape of width $\gamma$ decays in the spatial frequency domain over a range of $\Delta k_x \sim \gamma/2$ and can thus be taken as approximately band-limited. Applying the Nyquist-Shannon sampling theorem\cite{Shannon1984,Nyquist2002} then implies that to avoid under sampling, and hence informational oscillations, we require 
\begin{align}
\Delta \lesssim \frac{1}{2\Delta k_x} = \frac{\gamma}{4}.\label{eq:Nyquist}
\end{align}
Similar conclusions can also be made for alternative lineshapes. This criterion is indeed supported by the data shown in Figure~\ref{fig:pixels} which exhibits oscillations in the FI for the $N=100$ and $250$ cases, corresponding to $\gamma/\Delta = 3/5$ and $3/2$ respectively, whereas for $N=1000$ ($\gamma/\Delta = 6$) oscillations are negligible. The $N=500$ ($\gamma/\Delta = 3$) case lies close to the limit set by Eq.~\eqref{eq:Nyquist} such that oscillations may be expected, however, none are evident in Figure~\ref{fig:pixels}. Primarily, this behaviour is due to the fact that the approximations made in deriving Eq.~\eqref{eq:Nyquist}  begin to break down when $\Delta/\gamma \sim 1$. As such, Eq.~\eqref{eq:Nyquist} should only be viewed as a general experimental rule of thumb.

\subsection*{Poisson distributed noise}
\begin{figure}[t!]
	\begin{center}
		\includegraphics[width=\columnwidth]{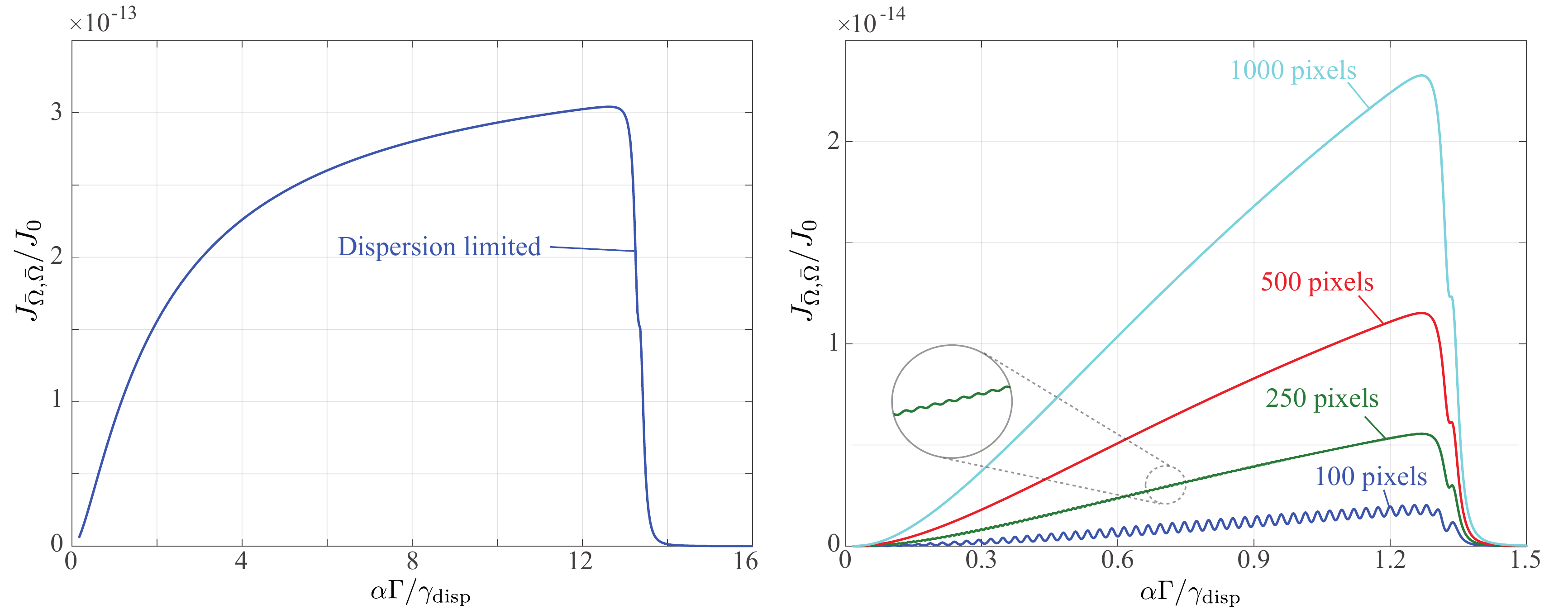}
		\caption{As per (left) Figure~\ref{fig:Lorentz_approx} and (right) Figure~\ref{fig:pixels} albeit assuming Poisson distributed  noise.  \label{fig:Poisson}}
	\end{center}
\end{figure}

Hitherto, discussion of the precision in inelastic optical spectroscopy has been limited to the simpler case of Gaussian distributed noise. This model is appropriate when read out noise dominates or if the mean signal is sufficiently high. In this section, however, we relax this assumption and instead briefly consider the effects of Poisson noise. To begin we must revisit the form of the FIM as was previously given by Eq.~\eqref{eq:FIMsum}. For Poisson distributed noise (again assuming the noise on each pixel is independent) it can be shown that $\mathbb{J}_{\mathbf{I}_{\mbox{\tiny{det}}}}$ is a diagonal matrix with on diagonal elements given by $1/I_j$\cite{Scharf1991}. Accordingly, the FIM is given by
\begin{align}
[\mathbb{J}_{\mathbf{w}}]_{kl} = \sum_{j}\! \frac{1}{I_j} \partialdiff{I_j}{w_k}\partialdiff{I_j}{w_l}. \label{eq:FIMsumPoisson}
\end{align}
The additional $1/I_j$ factor complicates the analysis of the limiting cases discussed above considerably. Particularly, we note that evaluation of the integrals involved (e.g. the analog of Eq.~\eqref{eq:FIM3}) can not be performed analytically. Numerical determination of the FIM in the general case can, however, be performed. In this vein, Figure~\ref{fig:Poisson}, shows the results of numerical calculations analogous to those presented in Figures~\ref{fig:Lorentz_approx} and \ref{fig:pixels}, albeit within a Poisson noise regime, i.e. using Eq.~\eqref{eq:FIMsumPoisson}. Note that for these plots we have defined $J_0 = N^2 \Delta \sum_{q = -\infty}^\infty |B_q|^4$. Two points of interest can be made based on Figure~\ref{fig:Poisson}. Firstly, in contrast to Figure~\ref{fig:Lorentz_approx}, no peak in the obtainable FI is seen. This difference arises because the signal to noise ratio from Gaussian and Poisson distributed noise exhibit different dependencies on the mean intensity recorded on a pixel. As such the balance between the improvement in signal to noise ratio, e.g. from a narrower instrumental peak width $\gamma_{\txtpow{disp}}$, and the effect of the linear scaling factor $\alpha$ discussed earlier is altered. In the Poisson noise regime the best precision is thus found to occur when the spectrum fills the detector without clipping of the spectral peaks. Secondly, the right hand plot of Figure~\ref{fig:Poisson} clearly shows that as the number of detector pixels is increased, the obtainable FI also increases. This trend similarly arises as a result of the decrease in the mean intensity recorded on each pixel as the number of pixels increases (for a fixed intensity distribution). Whilst for the Gaussian case the noise variance is fixed giving rise to a limiting FI as pixel count increases (as seen in Figure~\ref{fig:pixels}), in the Poisson case the drop in mean intensity implies that the noise variance also decreases such that the mean intensity can be determined more precisely. In the presence of Poisson noise, detectors with a finer pixelation thus not only enable greater estimation precision, but they also help avoid informational oscillations which can still be present at low pixel counts.

\section*{Conclusions}\label{sec:conclusions}
In this paper we have established the fundamental precision limits achievable when determining key spectral parameters in inelastic optical spectroscopy. In such applications, parameters such as the frequency shift of Raman or Brillouin scattered photons and the linewidth of the corresponding spectral peaks are of interest since they can provide quantitative information about the micro-mechanical, elastic and molecular information of material samples. Precision limits were derived using the concept of Fisher information and the Cram\'er-Rao lower bound taking into account the instrument response function of the dispersive element and potential defocus or spherical aberration in the detection optics, in addition to the size and pixelation of the detector itself. As such our results are applicable to a broad range of spectroscopic instruments. We also note that such limits are asymptotically achievable through use of a maximum likelihood estimation strategy. Whilst our analysis only employed a one-dimensional description of the lens, we note that this allowed significant physical insight to be gained through the analytic derivations it enabled. Moreover, a one-dimensional model is valid whenever one dimension of the PSF is large compared to the other. Since cylindrical lenses are frequently used in spectroscopic applications this is often the case. Nevertheless, extension of the above treatment to two dimensions is simple, albeit mathematically more involved. In particular, such a generalisation would only affect the optical model, for example, through use of Zernike aberration functions and two dimensional convolution integrals. Information theoretic aspects of our analysis would, however, remain unaffected except that the summations in Eqs.~\eqref{eq:FIMsum} and \eqref{eq:FIMsumPoisson} would be taken over a two dimensional array of detector pixels. Qualitatively similar behaviour to that found using the one-dimensional treatment, with only minor numerical differences, would be expected.

Although general formulae were given for the obtainable precision, a number of limiting cases were also considered providing greater insight into the obtainable precision. Specifically, we gave simplified results for both a dispersion and diffraction limited spectrometer. Optimal configurations (in terms of angular dispersion or optical magnification) could then be found for estimation of either the central frequency, inelastic shift or peak width in the Gaussian noise regime. These optima occur when the competing effects of the experimental lineshape and spatial separation of peaks on the detector are balanced. Numerical calculations were also presented for more general scenarios, including a discussion of the effects of Poisson noise. For fixed dispersion/magnification the obtainable precision was found to worsen as the widths of the response functions increased. Defocus and spherical aberration in the relay optics were also found to reduce the obtainable precision as would be intuitively expected, however, their effect on determining the optimal spectrometer configurations was found to be minimal even for relatively strong aberration strengths. Finally, although detector pixelation was shown to imply that informational limits can be highly sensitive to detector registration and other experimental parameters, this was only found to be significant in the coarse pixelation regime whereby spectral linewidths and pixel size are comparable, i.e. when the spectrum is sampled below the Nyquist rate. Given the high pixel count on many modern day detectors, this is unlikely to be of practical importance.

\section*{Acknowledgements}

This work was funded by the Royal Society through a University Research Fellowship. 

\section*{Author contributions statement}

 PT conceived the idea and performed aberration tolerance calculations. MRF performed analytic derivations and numerical calculations. Both authors wrote the article.

\section*{Additional information}

\textbf{Competing interests:} The authors declare no competing interests.

\end{document}